\documentclass[useAMS,usenatbib,usegraphicx]{mn2e}
\usepackage{amssymb}
\def\etal{\textit{et al.}\ }

\def\eg{e.\,g.\,}

\def\pmi{$\pm$}

\title[Polarization in young open cluster NGC 6823]{Polarization in young open cluster NGC 6823}
\author[B.J.Medhi et. al]
{Biman J. Medhi $^{1}$\thanks{E-mail: biman@aries.res.in},
Maheswar. G$^{1}$, J. C. Pandey$^{1}$, Motohide Tamura$^{2}$  and R. Sagar$^{1}$\\
$^{1}$Aryabhatta Research Institute of Observational Sciences (ARIES), Manora Peak, Nainital - 263 129, India \\
$^{2}$National Astronomical Observatory of Japan, Mitaka, Tokyo 181-8588, Japan \\}
\begin{document}

\date{}

\pagerange{\pageref{firstpage}--\pageref{lastpage}} \pubyear{2009}

\maketitle
\label{firstpage}
\begin{abstract}
We  present multiwavelength  linear polarimetric  observations  of 104
stars  towards  the  region  of  young open  cluster  NGC  6823.   The
polarization towards  NGC 6823 is dominated by  foreground dust grains
and we found  the evidence for the presence of  several layers of dust
towards  the  line of  sight.   The first  layer  of  dust is  located
approximately within 200 pc towards  the cluster, which is much closer
to the Sun than the cluster ($\sim 2.1$ kpc).  The radial distribution
of  the position  angles for  the  member stars  are found  to show  a
systematic change  while the polarization found to  reduce towards the
outer parts of  the cluster and the average  position angle of coronal
region of the cluster is very close to the inclination of the Galactic
parallel  ($\sim32^{\circ}$).   The size  distribution  of the  grains
within  NGC   6823  is  similar  to  those   in  general  interstellar
medium.  The  patchy  distribution   of  foreground  dust  grains  are
suggested to be mainly responsible for the both differential reddening
and polarization towards NGC 6823.  The majority of the observed stars
do not show the evidence of intrinsic polarization in their light.
\end{abstract}


\section{Introduction}

The  polarization   of  starlight  through   selective  extinction  by
aligned/partially  aligned  and  asymmetric  dust grains,  present  in
general  interstellar medium  can be  consider as  a valuable  tool to
study both the  grain properties like size and  shape, and small/large
scale structure  of interstellar magnetic fields in  different line of
sight. Although the identity of the dominant grain alignment mechanism
has proved  to be an  intriguing problem in grain  dynamics (Lazarian,
Goodman \& Myers  1997), it is generally believed  that the asymmetric
grains tend to  become aligned to the local  magnetic field with their
shortest  axis  parallel to  the  field.   For  this orientation,  the
observed   polarization  vector  is   parallel  to   the  plane-of-sky
projection  of  a  line-of-sight-averaged  magnetic  field  (Davis  \&
Greenstein  1951).   Interstellar  polarization strongly  varies  with
wavelength (Serkowski,  Mathewson \& Ford 1975; Wilking  , Lebofsky \&
Rieke 1982).   In particular,  the wavelength of  maximum interstellar
polarization  $(\lambda_{max})$  is  thought  to  be  related  to  the
total-to-selective   extinction   $(R_V)$   as   $R_{V}=(5.6\pm   0.3)
\lambda_{max}$(Whittet \&  van Breda 1978).  Generally, the unreddened
stars  show  no polarization,  if  it is  not  a  source of  intrinsic
polarization.  The stars with  large colour excess (E(B-V)) the values
of   polarization    P   show    a   wide   distribution    given   by
$P/E(B-V)_{max}=0.090\  mag^{-1}$,  where  P  is  measured  at  visual
wavelength (Spitzer 1978).

The polarimetric study of young  open clusters can provide us valuable
information  about   the  foreground  interstellar   dust  because  of
the available  knowledge  on  their  physical  parameters  like  distance,
membership  probability (Mp)  and  colour  excess.  As  a  part of  an
observational programme to carry  out the polarimetric observations of
young open clusters to  investigate the properties like magnetic field
orientation, $\lambda_{max}$ , maximum polarization $(P_{max})$, etc.,
(Medhi \etal 2007, 2008) we  observed the young open cluster NGC 6823,
which  is  Hubble  Space  Telescope's (HST)  polarimetric  calibration
object  (Turnshek  \etal, 1990).   The  young  open  cluster NGC  6823
$(R.A.(J2000):  19^{h}  43^{m}  09^{s}, Dec(J2000):+23^{\circ}  18^{m}
00^{s}; l=59.402,  b=-0.144)$ is  the central cluster  of the  Vul OBI
association  (Morgan \etal,  1953).  It  is located  in the  local arm
(Orion)  of our Galaxy.   A distance  modulus of  $11.6\pm 0.01$ mag
which corresponds to a distance of  $2.1 \pm 0.1$ kpc is estimated for
the cluster  (Guetter 1992).   The bright main  sequence stars  in NGC
6823 reveal  an age of $\sim 5$  Myr for the cluster,  whereas the pre
main sequence stars  indicate an age of younger  than $0.3$ Myr (Sagar
\& Joshi 1981).

To study  the interstellar polarization in different  direction of our
galaxy Hiltner (1956), Hall (1958)  and Serkowski (1965)  made the
polarimetric measurements of few bright  stars in the direction of NGC
6823.   The four  bright stars  observed  by Hiltner  (1956) and  Hall
(1958),  using  Corning 3385  (equivalent  to  V  filter) and  without
(clear)  filter  respectively, are  common  and  belongs  to O  and  B
spectral type.  Serkowski (1965) had observed 21 stars in V filter and
17  stars in B  filter including  3 bright  stars observed  by Hiltner
(1956) and  Hall (1958). However they made the  observations in survey
mode, therefore no firm conclusion could be drawn from their study.

In  this paper, we  present the  results of  polarimetric measurements
made for  104 stars in  $B, V$ and  $R_C$ photometric band  toward NGC
6823 (brighter than $V\simeq 17$  mag). Out of the 104 stars observed,
42 of them  have high membership probability ($M_{P}  \geq 0.5$).  The
paper is organized  as following: in section 2  we present observation
and  data  reduction; the  results  and  discussion  are presented  in
section 3 and in section 4 we conclude with a summary.
\begin{table*}
\centering
\caption{Observed polarized and unpolarized standard stars}\label{std_obs}
\begin{tabular}{lllllll}
\hline
\multicolumn{5}{|c|}{Polarized Standard}&\multicolumn{2}{c}{Unpolarized Standard}\\
\hline
Filter&$P\pm\epsilon(\%)$ &  $\theta \pm \epsilon(^\circ)$ &  $P\pm\epsilon(\%)$ & $\theta \pm\epsilon(^\circ)$ &\ \ \ \ $q(\%)$ &\ \ \ \ \ $u(\%)$ \\
\hline
& \multicolumn{2}{c}{Schmidt, Elston \& Lupie(1992)}&\multicolumn{2}{c}{This work}&\multicolumn{2}{c}{This work}\\
\hline
\multicolumn{5}{|c|}{\underline{Hiltner-960}}                            &\multicolumn{2}{|c|}{\underline{HD21447}} \\
B & $5.72\pm0.06$ & $55.06\pm 0.31$  & $5.69\pm 0.20$&$  55 \pm 2$ & \ \ \ \ \ 0.019  &\ \ \ \ \ \ 0.011  \\
V & $5.66\pm0.02$ & $54.79\pm 0.11$  & $5.61\pm 0.14$&$  53 \pm 2$ & \ \ \ \ \ 0.037  &\ \ \ \ -\ 0.031 \\
R & $5.21\pm0.03$ & $54.54\pm 0.16$  & $5.20\pm 0.06$&$  54 \pm 2$ & \ \ \ -\ 0.035  &\ \ \ \ -\ 0.039 \\
\multicolumn{5}{|c|}{\underline{HD 204827}}                              &\multicolumn{2}{|c|}{\underline{HD12021}} \\
B & $5.65\pm0.02$ & $58.20\pm 0.11$  & $5.72\pm 0.09$&$  59 \pm 2$ & \ \ \ -\ 0.081  & \ \ \ \ \ \ 0.071 \\
V & $5.32\pm0.02$ & $58.73\pm 0.08$  & $5.35\pm 0.03$&$  60 \pm 2$ & \ \ \ \ \ 0.042 & \ \ \ \ -\ 0.045 \\
R & $4.89\pm0.03$ & $59.10\pm 0.17$  & $4.90\pm 0.20$&$  59 \pm 2$ & \ \ \ \ \ 0.020 & \ \ \ \ \ \ 0.031 \\
\multicolumn{5}{|c|}{\underline{BD+64$^{\circ}$106}}                     &\multicolumn{2}{|c|}{\underline{HD14069}} \\
B & $5.51\pm0.09$ & $97.15\pm 0.47$  & $5.46\pm 0.10$&$  99 \pm 2$ &\ \ \ \ \ 0.038 &\ \ \ \ -\ 0.010 \\
V & $5.69\pm0.04$ & $96.63\pm 0.18$  & $5.58\pm 0.11$&$  97 \pm 2$ &\ \ \ \ \ 0.021 &\ \ \ \ \ \  0.018 \\
R & $5.15\pm0.10$ & $96.74\pm 0.54$  & $5.20\pm 0.02$&$  97 \pm 2$ &\ \ \ \ \ 0.010 &\ \ \ \ -\ 0.014 \\
\multicolumn{5}{|c|}{\underline{HD 19820}}                               &\multicolumn{2}{|c|}{\underline{G191B2B}} \\
B & $4.70\pm0.04$ & $115.70\pm 0.22$ & $4.75\pm 0.20$&$ 114 \pm 2$ &\ \ \ \ \ 0.072 &\ \ \ \ -\ 0.059 \\
V & $4.79\pm0.03$ & $114.93\pm 0.17$ & $4.81\pm 0.10$&$ 114 \pm 2$ &\ \ \ -\ 0.022 &\ \ \ \ -\ 0.041 \\
R & $4.53\pm0.03$ & $114.46\pm 0.17$ & $4.65\pm 0.13$&$ 113 \pm 2$ &\ \ \ -\ 0.036 &\ \ \ \ \ \ 0.027 \\
\hline
\multicolumn{7}{|c|}{Secondary Polarized Standard}\\
\hline
Star Name&Filter&$P\pm\epsilon(\%)$ &  $\theta $ &  $P\pm\epsilon(\%)$ & $\theta \pm\epsilon(^\circ)$ & \\
\hline
 & \multicolumn{3}{c}{\ \ \ \ \ \ \ \ Turnshek \etal (1990)}&\multicolumn{2}{c}{This work}&\\
\hline
NGC6823-2j&  V & $2.67\pm0.11$ & $13.0$  & $2.70\pm 0.01$&$ 11 \pm 2$  &   \\
NGC6823-6j&  V & $3.23\pm0.13$ & $09.0$  & $3.24\pm 0.07$&$ 08 \pm 2$  &   \\
NGC6823-7j&  V & $3.74\pm0.13$ & $06.9$  & $3.74\pm 0.15$&$ 06 \pm 2$  &   \\
NGC6823-10j& V & $4.55\pm0.13$ & $07.2$  & $4.57\pm 0.01$&$ 06 \pm 2$  &   \\
\hline

\end{tabular}
\label{std_obs}
\end{table*}

\begin{table*}			
\centering			
\begin{minipage}{650mm}		
\caption{Observed $B, V$ and $R_{c}$ polarization values for different stars in NGC 6823}\label{result_1}
\label{tab_2}
\begin{tabular}{lllllllllll}	
\hline	 			
\hline	 			
ID(M)$^\dag$& ID(B)$^\dag$$^\dag$ & V(mag) & $P_{B}$\pmi $\epsilon$ $(\%)$ & $\theta_{B}$ \pmi $\epsilon$ $(^{\circ})$ &  $P_{V}$ \pmi
$\epsilon$ $(\%)$ & $\theta_{V} $\pmi $\epsilon$ $(^{\circ})$& $P_{R}$\pmi  $\epsilon$ $(\%)$ & $\theta_{R} $
\pmi $\epsilon$ $(^{\circ})$&  $M_p(E)$$^*$ & $M_p(D)$$^*$$^*$ \\
(1)&(2)& (3)&(4)&(5)&(6)&(7)&(8)&(9)&(10)&(11)\\
\hline	 			
01  & --- &  16.06 &  1.93 \pmi 0.54 & 26  \pmi 9 &  2.15 \pmi 0.32  &  32  \pmi 5  &  1.88 \pmi 0.17  &  25  \pmi 4 &   ---  &  --- \\        
02  & 20  &  14.98 &  4.13 \pmi 0.16 & 2   \pmi 2 &  4.39 \pmi 0.09  &  2   \pmi 2  &  4.28 \pmi 0.11  &  4   \pmi 2 &   0.96 &  0.86 \\ 
03  & 74  &  13.77 &  2.56 \pmi 0.05 & 9   \pmi 2 &  2.70 \pmi 0.01  &  11  \pmi 2  &  2.55 \pmi 0.03  &  9   \pmi 2 &   0.97 &  0.72 \\ 
04  & 73  &  11.47 &  0.50 \pmi 0.07 & 23  \pmi 8 &  0.52 \pmi 0.09  &  25  \pmi 5  &  0.51 \pmi 0.01  &  22  \pmi 3 &   ---  &  ---  \\ 
05  & 18  &  16.36 &  3.47 \pmi 0.18 & 9   \pmi 2 &  3.34 \pmi 0.55  &  12  \pmi 5  &  3.54 \pmi 0.23  &  5   \pmi 2 &   ---  &  0.00 \\ 
06  & 23  &  16.29 &  2.32 \pmi 0.70 & 22  \pmi 5 &  2.51 \pmi 0.14  &  22  \pmi 2  &  2.47 \pmi 0.07  &  26  \pmi 2 &   ---  &  0.88 \\ 
07  & 17  &  14.2  &  3.72 \pmi 0.01 & 4   \pmi 2 &  3.74 \pmi 0.15  &  6   \pmi 2  &  3.73 \pmi 0.03  &  5   \pmi 2 &   0.97 &  0.77 \\ 
08  & 78  &  14.27 &  3.15 \pmi 0.03 & 2   \pmi 2 &  3.69 \pmi 0.07  &  4   \pmi 2  &  3.44 \pmi 0.08  &  2   \pmi 2 &   0.93 &  0.80 \\ 
09  & --- &  16.25 &  4.17 \pmi 0.21 & 4   \pmi 2 &  4.10 \pmi 0.30  &  5   \pmi 2  &  4.12 \pmi 0.08  &  3   \pmi 2 &   ---  &  0.83 \\ 
10  & 27  &  15.05 &  1.97 \pmi 0.01 & 23  \pmi 2 &  2.18 \pmi 0.14  &  27  \pmi 2  &  2.11 \pmi 0.15  &  22  \pmi 3 &   ---  &  0.01 \\ 
11  & 26  &  13.86 &  1.99 \pmi 0.07 & 18  \pmi 2 &  1.98 \pmi 0.06  &  20  \pmi 2  &  1.88 \pmi 0.03  &  16  \pmi 2 &   ---  &  0.00 \\ 
12  & 13  &  11.98 &  3.06 \pmi 0.09 & 8   \pmi 2 &  3.13 \pmi 0.06  &  12  \pmi 2  &  3.06 \pmi 0.04  &  8   \pmi 2 &   0.95 &  0.81 \\ 
13  & 28  &  16.4  &  3.42 \pmi 0.52 & 3   \pmi 4 &  3.78 \pmi 0.49  &  5   \pmi 4  &  3.79 \pmi 0.04  &  8   \pmi 2 &   ---  &  0.88 \\ 
14  & 75  &  15.19 &  2.13 \pmi 0.16 & 23  \pmi 2 &  2.15 \pmi 0.17  &  23  \pmi 3  &  2.20 \pmi 0.05  &  24  \pmi 2 &   ---  &  0.88 \\ 
15  & 77  &  15.13 &  1.63 \pmi 0.03 & 21  \pmi 2 &  1.72 \pmi 0.03  &  27  \pmi 2  &  1.66 \pmi 0.29  &  33  \pmi 6 &   ---  &  0.45 \\ 
16  & 76  &  15.78 &  3.12 \pmi 0.15 & 5   \pmi 2 &  3.24 \pmi 0.07  &  8   \pmi 2  &  3.19 \pmi 0.01  &  5   \pmi 2 &   ---  &  0.84 \\ 
17  & 16  &  16.06 &  3.15 \pmi 0.11 & 3   \pmi 2 &  3.73 \pmi 0.31  &  5   \pmi 2  &  3.08 \pmi 0.28  &  4   \pmi 3 &   ---  &  ---  \\ 
18  & 71  &  15.67 &  0.98 \pmi 0.18 & 26  \pmi 7 &  1.24 \pmi 0.10  &  17  \pmi 3  &  1.01 \pmi 0.08  &  17  \pmi 3 &   ---  &  0.34 \\ 
19  & 70  &  16.49 &  4.01 \pmi 0.29 & 3   \pmi 2 &  4.09 \pmi 0.16  &  4   \pmi 2  &  3.85 \pmi 0.41  &  7   \pmi 3 &   ---  &  0.89 \\ 
20  & 68  &  13.93 &  4.50 \pmi 0.06 & 4   \pmi 2 &  4.57 \pmi 0.01  &  6   \pmi 2  &  4.48 \pmi 0.20  &  4   \pmi 2 &   0.96 &  0.81 \\ 
21  & 12  &  16.26 &  0.62 \pmi 0.03 & 168 \pmi 3 &  0.80 \pmi 0.17  &  174 \pmi 4  &  0.66 \pmi 0.13  &  179 \pmi 3 &   ---  &  0.00 \\ 
22  & 11  &  13.54 &  5.04 \pmi 0.08 & 3   \pmi 2 &  5.28 \pmi 0.05  &  4   \pmi 2  &  4.94 \pmi 0.03  &  2   \pmi 2 &   0.97 &  0.82 \\ 
23  & 4   &  13.63 &  0.52 \pmi 0.05 & 23  \pmi 9 &  0.55 \pmi 0.04  &  23  \pmi 9  &  0.53 \pmi 0.05  &  24  \pmi 8 &   ---  &  0.00 \\ 
24  & 9   &  13.79 &  4.46 \pmi 0.04 & 12  \pmi 2 &  4.40 \pmi 0.03  &  13  \pmi 2  &  4.22 \pmi 0.13  &  12  \pmi 2 &   0.97 &  0.77 \\ 
25  & 8   &  15.82 &  1.79 \pmi 0.19 & 29  \pmi 3 &  1.83 \pmi 0.06  &  19  \pmi 2  &  1.81 \pmi 0.09  &  26  \pmi 2 &   ---  &  0.82 \\ 
26  & 5   &  14.88 &  1.49 \pmi 0.20 & 26  \pmi 4 &  1.82 \pmi 0.15  &  35  \pmi 3  &  1.41 \pmi 0.21  &  40  \pmi 6 &   ---  &  0.72 \\ 
27  & 64a &  14.5  &  3.33 \pmi 0.37 & 96  \pmi 3 &  3.39 \pmi 0.32  &  98  \pmi 3  &  3.25 \pmi 0.07  &  94  \pmi 2 &   ---  &  0.00 \\ 
28  & 65  &  15.56 &  5.21 \pmi 0.51 & 7   \pmi 3 &  5.07 \pmi 0.17  &  4   \pmi 2  &  4.97 \pmi 0.10  &  2   \pmi 2 &   ---  &  0.76 \\ 
29  & 64  &  15.07 &  0.86 \pmi 0.23 & 21  \pmi 8 &  0.83 \pmi 0.10  &  18  \pmi 7  &  0.80 \pmi 0.05  &  20  \pmi 4 &   ---  &  0.00 \\ 
30  & 62  &  14.63 &  1.34 \pmi 0.24 & 26  \pmi 6 &  1.44 \pmi 0.27  &  28  \pmi 7  &  1.39 \pmi 0.31  &  23  \pmi 8 &   0.95 &  0.82 \\ 
31  & 61  &  15.49 &  0.95 \pmi 0.13 & 27  \pmi 7 &  0.94 \pmi 0.16  &  21  \pmi 4  &  0.91 \pmi 0.18  &  21  \pmi 7 &   ---  &  0.41 \\ 
32  & 63  &  16.31 &  1.40 \pmi 0.22 & 15  \pmi 6 &  1.42 \pmi 0.23  &  20  \pmi 9  &  1.44 \pmi 0.29  &  24  \pmi 8 &   ---  &  0.59 \\ 
33  & 65a &  16.09 &  1.19 \pmi 0.24 & 17  \pmi 7 &  1.22 \pmi 0.28  &  17  \pmi 9  &  1.27 \pmi 0.22  &  16  \pmi 7 &   ---  &  0.77 \\ 
34  & --- &  16.51 &  0.94 \pmi 0.18 & 35  \pmi 7 &  1.23 \pmi 0.21  &  42  \pmi 7  &  0.99 \pmi 0.17  &  32  \pmi 8 &   ---  &  0.87 \\ 
35  & 25  &  16.48 &  3.68 \pmi 0.22 & 2   \pmi 3 &  3.73 \pmi 0.11  &  6   \pmi 2  &  3.99 \pmi 0.26  &  4   \pmi 6 &   ---  &  0.89 \\ 
36  & --- &  16.44 &  3.03 \pmi 0.25 & 8   \pmi 2 &  3.03 \pmi 0.39  &  7   \pmi 4  &  2.83 \pmi 0.31  &  4   \pmi 3 &   ---  &  ---  \\ 
37  & 10  &  16.44 &  5.94 \pmi 0.18 & 9   \pmi 2 &  5.50 \pmi 0.93  &  9   \pmi 6  &  5.55 \pmi 0.18  &  12  \pmi 2 &   ---  &  0.48 \\ 
38  & --- &  16.71 &  3.47 \pmi 0.23 & 12  \pmi 2 &  3.71 \pmi 0.65  &  16  \pmi 5  &  3.41 \pmi 0.22  &  18  \pmi 2 &   ---  &  0.45 \\ 
39  & --- &  16.43 &  4.88 \pmi 0.09 & 15  \pmi 2 &  4.92 \pmi 0.15  &  13  \pmi 2  &  4.83 \pmi 0.06  &  12  \pmi 2 &   ---  &  0.89 \\ 
40  & 14  &  14.17 &  2.88 \pmi 0.06 & 2   \pmi 2 &  3.31 \pmi 0.05  &  3   \pmi 2  &  2.97 \pmi 0.06  &  3   \pmi 2 &   0.97 &  0.85 \\ 
41  & 33  &  14.99 &  0.99 \pmi 0.14 & 16  \pmi 9 &  1.06 \pmi 0.18  &  18  \pmi 8  &  1.01 \pmi 0.13  &  11  \pmi 5 &   0.82 &  0.83 \\ 
42  & 36  &  15.71 &  2.26 \pmi 0.21 & 25  \pmi 4 &  2.42 \pmi 0.23  &  31  \pmi 3  &  2.35 \pmi 0.05  &  27  \pmi 2 &   ---  &  0.57 \\ 
43  & --- &  16.77 &  2.29 \pmi 0.67 & 30  \pmi 8 &  2.34 \pmi 0.54  &  32  \pmi 7  &  2.27 \pmi 0.42  &  35  \pmi 5 &   ---  &  ---  \\ 
44  & 34  &  13.91 &  3.52 \pmi 0.10 & 9   \pmi 2 &  3.65 \pmi 0.16  &  10  \pmi 2  &  3.55 \pmi 0.03  &  10  \pmi 2 &   0.98 &  0.59 \\ 
45  & 51  &  16.06 &  1.91 \pmi 0.20 & 11  \pmi 4 &  2.10 \pmi 0.11  &  18  \pmi 2  &  1.96 \pmi 0.17  &  16  \pmi 2 &   ---  &  0.01 \\ 
46  & 53  &  16.09 &  1.79 \pmi 0.30 & 15  \pmi 6 &  1.85 \pmi 0.28  &  19  \pmi 5  &  1.76 \pmi 0.06  &  18  \pmi 2 &   ---  &  0.02 \\ 
47  & 35  &  11.37 &  1.52 \pmi 0.02 & 29  \pmi 2 &  1.63 \pmi 0.09  &  29  \pmi 2  &  1.47 \pmi 0.10  &  28  \pmi 2 &   0.79 &  ---  \\ 
48  & 50  &  14.32 &  0.55 \pmi 0.02 & 42  \pmi 2 &  0.62 \pmi 0.08  &  44  \pmi 9  &  0.58 \pmi 0.06  &  38  \pmi 7 &   ---  &  ---  \\ 
49  & 49  &  15.49 &  1.16 \pmi 0.13 & 17  \pmi 4 &  1.09 \pmi 0.06  &  18  \pmi 2  &  1.09 \pmi 0.10  &  11  \pmi 4 &   ---  &  ---  \\ 
50  & --- &  16.48 &  1.65 \pmi 0.36 & 26  \pmi 7 &  1.76 \pmi 0.36  &  33  \pmi 7  &  1.78 \pmi 0.22  &  35  \pmi 4 &   ---  &  ---  \\ 
51  & 42  &  13.92 &  0.88 \pmi 0.16 & 12  \pmi 9 &  0.98 \pmi 0.01  &  15  \pmi 2  &  0.95 \pmi 0.01  &  13  \pmi 2 &   ---  &  ---  \\ 
52  & 98  &  14.38 &  1.35 \pmi 0.39 & 17  \pmi 8 &  1.44 \pmi 0.31  &  19  \pmi 5  &  1.47 \pmi 0.32  &  17  \pmi 3 &   ---  &  0.00 \\ 
53  & 99  &  16.01 &  2.52 \pmi 0.47 & 24  \pmi 4 &  2.75 \pmi 0.28  &  26  \pmi 3  &  2.46 \pmi 0.14  &  22  \pmi 4 &   ---  &  0.00 \\ 
54  & 80  &  14.27 &  1.75 \pmi 0.07 & 17  \pmi 2 &  2.15 \pmi 0.25  &  20  \pmi 4  &  1.83 \pmi 0.12  &  18  \pmi 4 &   ---  &  0.69 \\ 
55  & 79  &  15.37 &  3.02 \pmi 0.15 & 21  \pmi 2 &  3.20 \pmi 0.64  &  21  \pmi 2  &  3.11 \pmi 0.66  &  27  \pmi 2 &   ---  &  0.82 \\ 
56  & --- &  15.57 &  3.16 \pmi 0.18 & 10  \pmi 2 &  3.39 \pmi 0.52  &  13  \pmi 5  &  3.26 \pmi 0.33  &  11  \pmi 2 &   ---  &  0.02 \\ 
57  & 81  &  14.32 &  4.02 \pmi 0.21 & 172 \pmi 2 &  4.54 \pmi 0.11  &  176 \pmi 2  &  3.93 \pmi 0.15  &  170 \pmi 5 &   ---  &  ---  \\ 
58  & --- &  16.70 &  1.96 \pmi 0.37 & 28  \pmi 9 &  2.26 \pmi 0.27  &  29  \pmi 4  &  2.01 \pmi 0.26  &  24  \pmi 4 &   ---  &  0.73 \\ 
59  & 82  &  13.29 &  0.87 \pmi 0.13 & 19  \pmi 6 &  0.91 \pmi 0.08  &  19  \pmi 5  &  0.85 \pmi 0.04  &  18  \pmi 4 &   0.55 &  ---  \\ 
60  & 125 &  15.01 &  0.61 \pmi 0.10 & 24  \pmi 8 &  0.79 \pmi 0.08  &  28  \pmi 6  &  0.68 \pmi 0.04  &  26  \pmi 5 &   ---  &  ---  \\      
61  & --- &  15.43 &  2.35 \pmi 0.25 & 16  \pmi 9 &  2.89 \pmi 0.20  &  21  \pmi 6  &  2.65 \pmi 0.22  &  17  \pmi 2 &   ---  &  0.29 \\      
62  & --- &  12.19 &  0.59 \pmi 0.04 & 37  \pmi 4 &  0.67 \pmi 0.07  &  38  \pmi 6  &  0.63 \pmi 0.11  &  34  \pmi 4 &   ---  &  ---  \\      
\hline																		  
\end{tabular}																	  
\end{minipage}																	  
\end{table*}																	  
\begin{table*}			
\centering			
\begin{minipage}{650mm}		
\caption{Continuation of Table 2}
\label{tab_3}
\begin{tabular}{lllllllllll}	
\hline	 			
\hline	 			
ID(M)$^\dag$ &  ID(B)$^\dag$$^\dag$ &V(mag) & $P_{B}$\pmi $\epsilon$ $(\%)$ & $\theta_{B}$ \pmi $\epsilon$ $(^{\circ})$ &  $P_{V}$ \pmi
$\epsilon$ $(\%)$ & $\theta_{V} $\pmi $\epsilon$ $(^{\circ})$& $P_{R}$\pmi  $\epsilon$ $(\%)$ & $\theta_{R} $
\pmi $\epsilon$ $(^{\circ})$& $M_p(E)$$^*$& $M_p(D)$$^*$$^*$ \\
(1)&(2)& (3)&(4)&(5)&(6)&(7)&(8)&(9)&(10)&(11)\\
\hline	 			
 63  & 96  &  15.06 &  1.73 \pmi 0.39 & 12  \pmi 3 &  1.68 \pmi 0.33  &  13  \pmi 4  &  1.64 \pmi 0.20  &  10  \pmi 3 &  ---  & 0.00\\    	      
 64  & --- &  15.82 &  2.85 \pmi 0.42 & 6   \pmi 5 &  3.20 \pmi 0.45  &  9   \pmi 4  &  2.79 \pmi 0.14  &  7   \pmi 5 &  ---  & --- \\    	      
 65  & --- &  16.29 &  4.80 \pmi 0.34 & 41  \pmi 3 &  4.86 \pmi 0.34  &  38  \pmi 3  &  4.77 \pmi 0.32  &  36  \pmi 4 &  ---  & --- \\    	      
 66  & --- &  15.08 &  1.65 \pmi 0.45 & 8   \pmi 9 &  1.63 \pmi 0.14  &  13  \pmi 3  &  1.55 \pmi 0.23  &  12  \pmi 3 &  ---  & --- \\    	      
 67  & 88  &  14.12 &  0.65 \pmi 0.10 & 6   \pmi 5 &  0.72 \pmi 0.05  &  8   \pmi 4  &  0.66 \pmi 0.04  &  7   \pmi 4 &  ---  & 0.40\\    	      
 68  & --- &  15.65 &  1.35 \pmi 0.20 & 36  \pmi 3 &  1.39 \pmi 0.17  &  37  \pmi 2  &  1.27 \pmi 0.17  &  36  \pmi 3 &  ---  & 0.58\\    	      
 69  & 86  &  14.50 &  0.95 \pmi 0.30 & 23  \pmi 6 &  1.21 \pmi 0.18  &  25  \pmi 6  &  1.01 \pmi 0.11  &  24  \pmi 6 &  ---  & --- \\    	      
 70  & 127 &  15.41 &  3.04 \pmi 0.06 & 10  \pmi 2 &  3.42 \pmi 0.06  &  18  \pmi 2  &  3.32 \pmi 0.03  &  11  \pmi 2 &  ---  & --- \\    	      
 71  & 130 &  14.83 &  1.11 \pmi 0.13 & 23  \pmi 6 &  1.21 \pmi 0.07  &  32  \pmi 2  &  1.25 \pmi 0.23  &  28  \pmi 8 &  ---  & --- \\    	      
 72  & 129 &  14.66 &  3.96 \pmi 0.27 & 15  \pmi 2 &  4.20 \pmi 0.36  &  14  \pmi 2  &  3.76 \pmi 0.06  &  17  \pmi 2 &  ---  & --- \\    	      
 73  & 95  &  14.38 &  1.32 \pmi 0.13 & 33  \pmi 3 &  1.45 \pmi 0.08  &  35  \pmi 2  &  1.37 \pmi 0.05  &  31  \pmi 2 &  0.75 & --- \\       
 74  & 93  &  12.08 &  2.69 \pmi 0.01 & 6   \pmi 2 &  2.74 \pmi 0.04  &  4   \pmi 2  &  2.52 \pmi 0.09  &  2   \pmi 2 &  0.97 & 0.81\\      
 75  & 94  &  14.50 &  2.82 \pmi 0.55 & 10  \pmi 6 &  3.16 \pmi 0.44  &  17  \pmi 4  &  2.79 \pmi 0.12  &  9   \pmi 2 &  ---  & --- \\    	      
 76  & 91  &  12.27 &  3.01 \pmi 0.03 & 11  \pmi 2 &  3.23 \pmi 0.08  &  12  \pmi 2  &  3.06 \pmi 0.10  &  13  \pmi 2 &  0.95 & --- \\       
 77  & --- &  15.33 &  1.27 \pmi 0.08 & 166 \pmi 2 &  1.26 \pmi 0.09  &  165 \pmi 3  &  1.18 \pmi 0.14  &  161 \pmi 5 &  ---  & --- \\    	      
 78  & --- &  15.78 &  1.50 \pmi 0.29 & 22  \pmi 3 &  1.66 \pmi 0.23  &  24  \pmi 3  &  1.43 \pmi 0.08  &  16  \pmi 2 &  ---  & --- \\    	      
 79  & --- &  16.19 &  1.96 \pmi 0.19 & 5   \pmi 3 &  2.31 \pmi 0.24  &  10  \pmi 3  &  2.02 \pmi 0.39  &  8   \pmi 5 &  ---  & --- \\    	      
 80  & --- &  16.3  &  3.23 \pmi 0.22 & 21  \pmi 2 &  3.37 \pmi 0.40  &  24  \pmi 4  &  3.46 \pmi 0.35  &  29  \pmi 6 &  ---  & 0.89\\    	      
 81  & 128 &  16.26 &  4.02 \pmi 0.29 & 6   \pmi 6 &  4.12 \pmi 0.28  &  8   \pmi 5  &  4.15 \pmi 0.12  &  9   \pmi 3 &  ---  & --- \\    	      
 82  & --- &  14.54 &  0.89 \pmi 0.22 & 7   \pmi 4 &  1.01 \pmi 0.20  &  8   \pmi 3  &  0.85 \pmi 0.21  &  4   \pmi 4 &  ---  & 0.46\\
 83  & --- &  14.35 &  1.29 \pmi 0.12 & 21  \pmi 3 &  1.51 \pmi 0.11  &  25  \pmi 2  &  1.15 \pmi 0.19  &  19  \pmi 3 &  ---  & --- \\    	      
 84  & --- &  14.29 &  1.16 \pmi 0.09 & 12  \pmi 3 &  1.21 \pmi 0.16  &  18  \pmi 6  &  1.07 \pmi 0.14  &  10  \pmi 5 &  ---  & --- \\    	      
 85  & --- &  11.78 &  0.73 \pmi 0.02 & 24  \pmi 3 &  0.76 \pmi 0.03  &  21  \pmi 2  &  0.77 \pmi 0.05  &  27  \pmi 3 &  ---  & --- \\    	      
 86  & --- &  15.99 &  2.63 \pmi 0.50 & 110 \pmi 7 &  2.78 \pmi 0.58  &  113 \pmi 8  &  2.79 \pmi 0.62  &  116 \pmi 8 &  ---  & --- \\    	      
 87  & --- &  15.44 &  3.58 \pmi 0.21 & 20  \pmi 2 &  3.67 \pmi 0.10  &  27  \pmi 2  &  3.48 \pmi 0.06  &  25  \pmi 2 &  ---  & --- \\    	      
 88  & --- &  13.86 &  2.35 \pmi 0.28 & 30  \pmi 4 &  2.46 \pmi 0.11  &  39  \pmi 2  &  2.34 \pmi 0.12  &  33  \pmi 2 &  ---  & --- \\           
 89  & --- &  15.64 &  3.11 \pmi 0.32 & 12  \pmi 6 &  3.41 \pmi 0.28  &  15  \pmi 4  &  3.26 \pmi 0.43  &  18  \pmi 8 &  ---  & --- \\    	      
 90  & --- &  15.81 &  4.17 \pmi 0.68 & 30  \pmi 6 &  4.23 \pmi 0.35  &  37  \pmi 2  &  4.26 \pmi 0.12  &  37  \pmi 2 &  ---  & --- \\    	      
 91  & --- &  16.21 &  3.18 \pmi 0.27 & 21  \pmi 3 &  3.24 \pmi 0.12  &  24  \pmi 2  &  3.16 \pmi 0.48  &  20  \pmi 5 &  ---  & --- \\    	      
 92  & 184 &  13.82 &  3.21 \pmi 0.11 & 24  \pmi 2 &  3.16 \pmi 0.21  &  21  \pmi 2  &  3.17 \pmi 0.28  &  24  \pmi 2 &  ---  & --- \\    	      
 93  & 179 &  13.12 &  1.85 \pmi 0.20 & 17  \pmi 3 &  2.05 \pmi 0.17  &  21  \pmi 2  &  1.87 \pmi 0.05  &  19  \pmi 2 &  0.97 & 0.89\\       
 94  & 180 &  12.5  &  2.49 \pmi 0.02 & 15  \pmi 2 &  2.75 \pmi 0.17  &  16  \pmi 2  &  2.41 \pmi 0.02  &  16  \pmi 2 &  ---  & 0.00\\   	      
 95  & --- &  15.05 &  2.45 \pmi 0.44 & 27  \pmi 5 &  2.60 \pmi 0.29  &  25  \pmi 3  &  2.35 \pmi 0.20  &  21  \pmi 3 &  ---  & --- \\    	      
 96  & 191 &  14.3  &  2.54 \pmi 0.21 & 19  \pmi 3 &  2.87 \pmi 0.44  &  18  \pmi 8  &  2.48 \pmi 0.27  &  22  \pmi 3 &  ---  & --- \\    	      
 97  & --- &  14.38 &  2.17 \pmi 0.59 & 18  \pmi 8 &  2.27 \pmi 0.09  &  25  \pmi 2  &  2.27 \pmi 0.39  &  28  \pmi 5 &  ---  & --- \\    	      
 98  & 185 &  14.48 &  2.78 \pmi 0.29 & 18  \pmi 4 &  2.89 \pmi 0.26  &  23  \pmi 2  &  2.69 \pmi 0.31  &  18  \pmi 3 &  ---  & 0.88\\    	      
 99  & 22  &  14.88 &  2.20 \pmi 0.43 & 36  \pmi 6 &  2.27 \pmi 0.36  &  38  \pmi 5  &  2.16 \pmi 0.11  &  37  \pmi 4 &  ---  & 0.07\\    	      
100  & 186 &  14.67 &  3.51 \pmi 0.17 & 10  \pmi 2 &  3.77 \pmi 0.03  &  13  \pmi 2  &  3.60 \pmi 0.16  &  13  \pmi 2 &  ---  & 0.57\\    	      
101  & 192 &  14.96 &  2.73 \pmi 0.10 & 35  \pmi 2 &  3.04 \pmi 0.37  &  27  \pmi 3  &  2.55 \pmi 0.23  &  21  \pmi 3 &  ---  & 0.47\\    	      
102  & 187 &  14.37 &  2.97 \pmi 0.05 & 20  \pmi 2 &  3.29 \pmi 0.02  &  18  \pmi 2  &  2.75 \pmi 0.10  &  20  \pmi 2 &  ---  & 0.81\\    	      
103  & --- &  14.63 &  4.11 \pmi 0.93 & 13  \pmi 2 &  4.20 \pmi 0.46  &  15  \pmi 3  &  3.95 \pmi 0.52  &  15  \pmi 4 &  ---  & 0.89\\    	      
104  & 29  &  15.33 &  4.30 \pmi 0.06 & 11  \pmi 2 &  4.33 \pmi 0.06  &  12  \pmi 2  &  4.35 \pmi 0.25  &  10  \pmi 2 &  ---  & 0.24\\    	      
\hline																		  
\end{tabular}																	  
\end{minipage}
\begin{quote}
{\hspace{.23cm}{ $^\dag$ : According to this observation}} \\
{\hspace{.176cm}{  $^\dag$ $^\dag$ : According to  Barkhatova (1957) }} \\
{\hspace{.5cm}{$^*$ : According to Erickson (1971)}}\\
{\hspace{.35cm}{$^*$$^*$\ : According to Dias \etal (2006)}}
\end{quote}																  
\end{table*}																	  
\begin{figure*}
\resizebox{12cm}{12cm}{\includegraphics{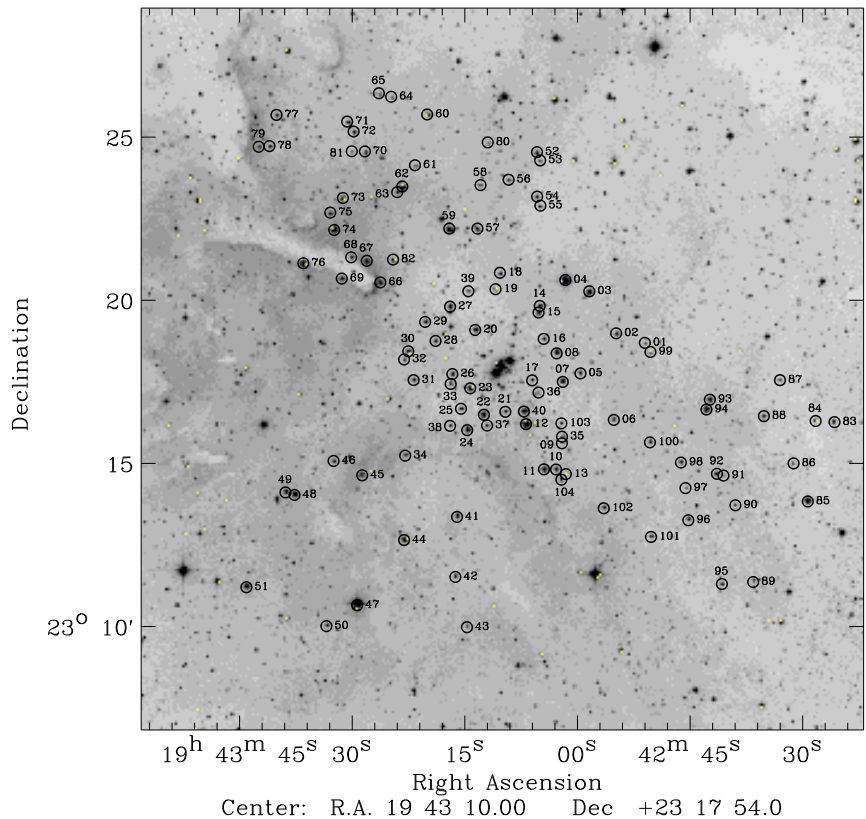}}
\caption{The  $25^{\prime}\times25^{\prime}$ R-band  DSS2 image  of the
field  containing NGC  6823, reproduced  from Digitized  Sky Survey II. The
field identification is followed by this observation (ID(M)).}
\label{NGC6823_field.ps}
\end{figure*}

\begin{figure*}
\resizebox{12cm}{12cm}{\includegraphics{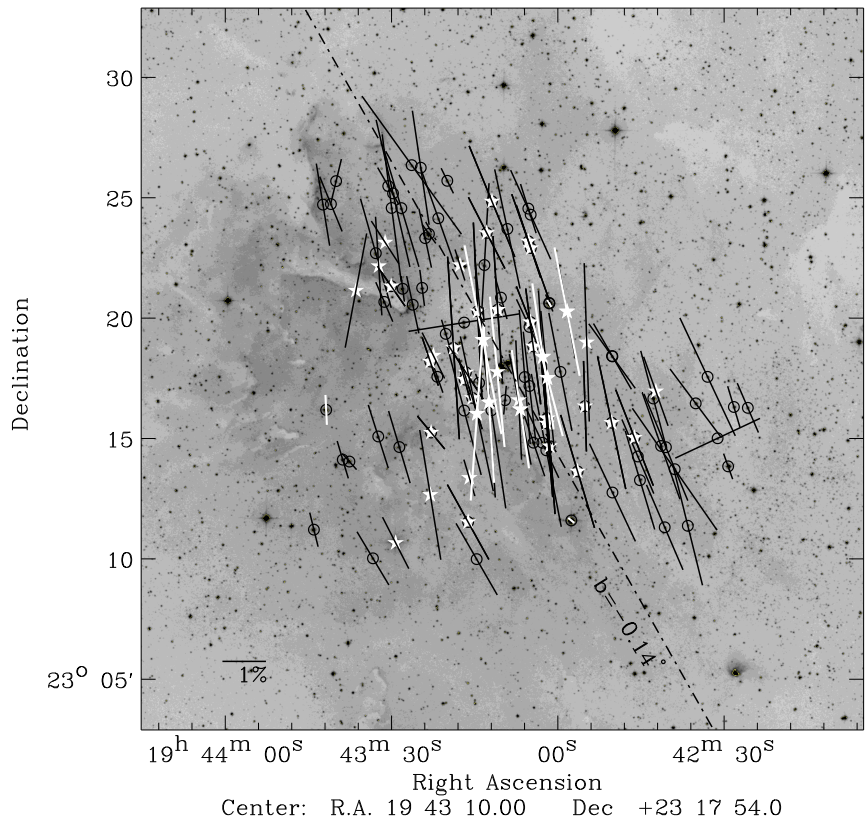}}
\caption{The  $30^{\prime}\times30^{\prime}$ R-band  DSS2 image  of the
  field  containing NGC  6823, reproduced  from Digitized  Sky Survey II.
  The  position  angles,  in  the equatorial  coordinate  system,  are
  measured  from  the  north,  increasing eastward.  The  polarization
  vectors  are  drawn with  the  star as  the  center.  Length of  the
  polarization   vector   is  proportional   to   the  percentage   of
  polarization $P_{V}$  and it is  oriented parallel to  the direction
  corresponding   to   the   observed  polarization   position   angle
  $\theta_{V}$. A vector  with a $P$ of $1\%$  is shown for reference.
  The    dashed   line   represents    the   Galactic    parallel   at
  $b=-0.14^{\circ}$.  Stars  with $M_{P}\geq0.50$ are  identified with
  closed star symbols in  white colour.  Polarization vectors of twelve
  stars observed by Serkowski et al.(1965) are shown in white colour.}
\label{NGC6823_V_dss.ps}
\end{figure*}

\begin{figure*}
\resizebox{12cm}{12cm}{\includegraphics{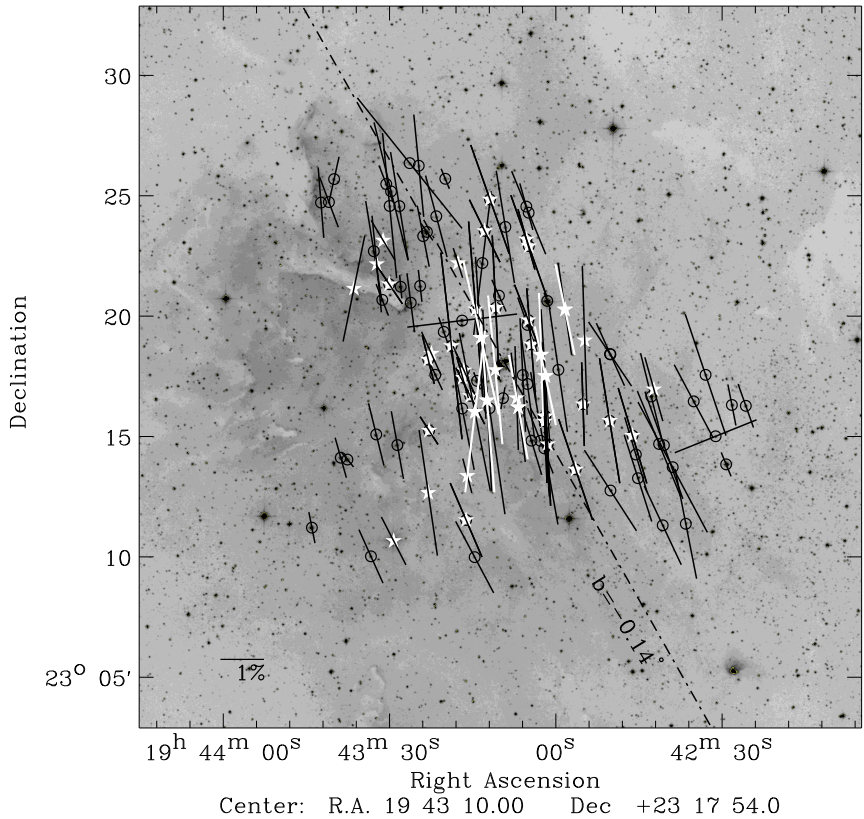}}
\caption{Same as in  Figure  \ref{NGC6823_V_dss.ps} but  for $P_{B}$ and
$\theta_{B}$.  Results for 9  stars observed  by Serkowski (1965) using B
filter are  shown using vectors drawn  in white.}
\label{NGC6823_B_dss.ps}
\end{figure*}

\begin{figure*}
\resizebox{12.0cm}{12.0cm}{\includegraphics{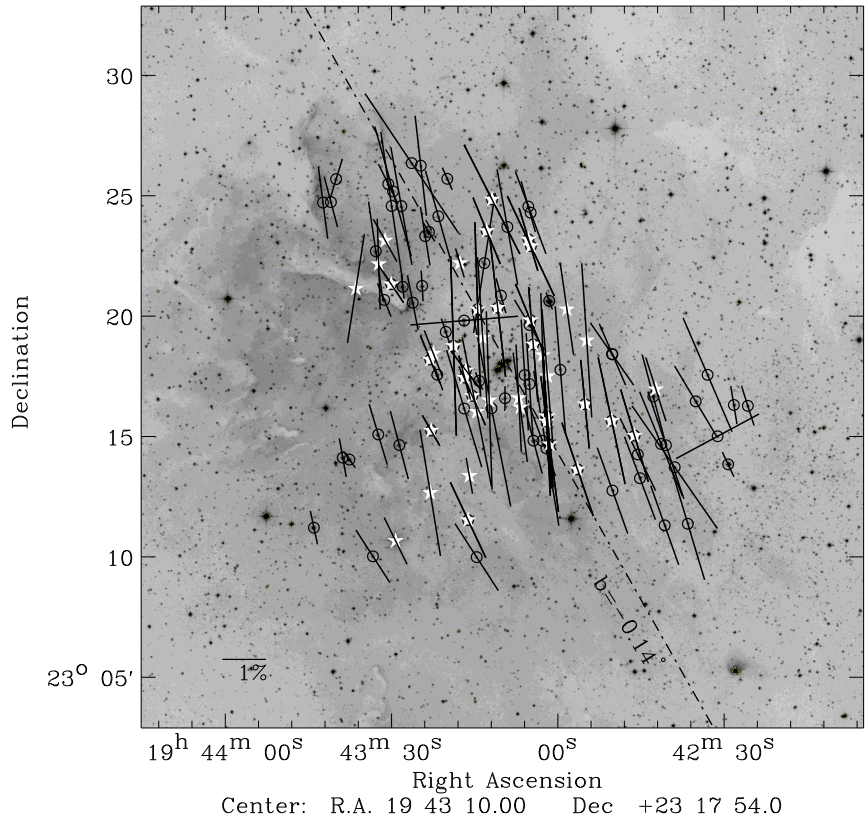}}
\caption{Same as in  Figure  \ref{NGC6823_V_dss.ps} but  for $P_{R}$ and
$\theta_{R}$.}
\label{NGC6823_R_dss.ps}
\end{figure*}

\begin{figure*}
\resizebox{12cm}{12.0cm}{\includegraphics{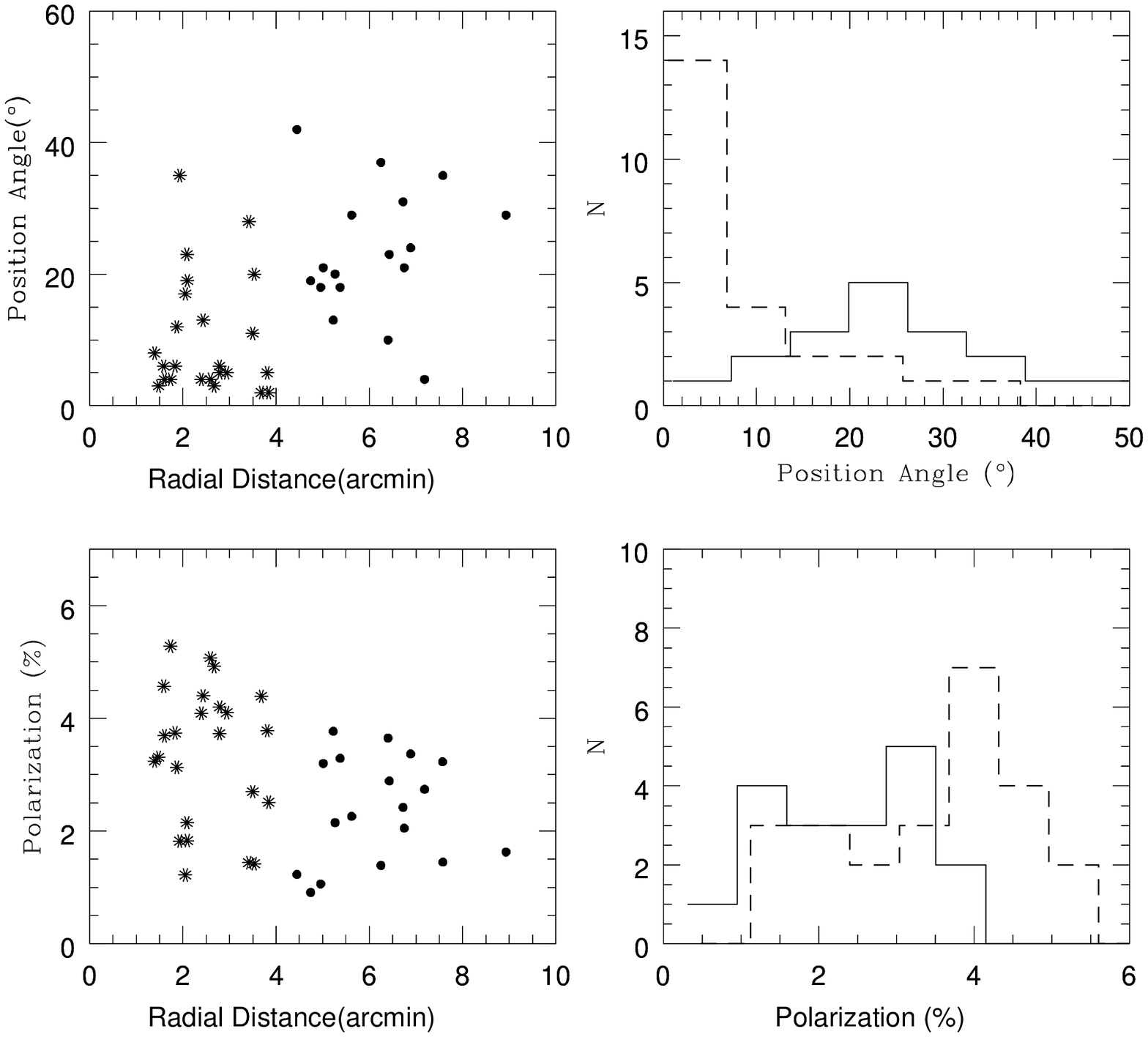}}
\caption{Distribution of $P_V$ and $PA_V$ for the 42 member stars with 
the radial distance from the cluster center}
\label{dif_pa.ps}
\end{figure*}

\section{Observation and Data Reduction}

The   polarimetric   data   for   the   four   fields   (centered   at
$R.A.:19^{h}43^{m}10^{s},\ Dec:+23^{\circ}17^{m}55^{s};\ R.A.:19^{h}43^{m}24^{s},
\    Dec:+23^{\circ}12^{m}26^{s};\    R.A.:19^{h}43^{m}24^{s},\   Dec:
+23^{\circ}23^{m}55^{s}$         and        $        R.A.:19^{h}42^{m}
4^{s},\ Dec:+23^{\circ}17^{m}43^{s}$) in  NGC 6823 were acquired using
ARIES Imaging Polarimeter  (AIMPOL; Medhi \etal 2007 )  mounted on the
Cassegrain  focus  of  the  104-cm Sampurnanand  Telescope  of  ARIES,
Nainital in $B$, $V$ and  $R_{c}$ photometric bands, on 24$^{th}$ May,
25$^{th}$ May, 06$^{th}$ June and 07$^{th}$ June 2007. The imaging was
done by using a TK $1024\times 1024$ pixel$^2$ CCD camera.  Each pixel
of the CCD corresponds to $1.7$  arcsec and the field of view is $\sim
8$ arcmin diameter  on the sky.  The FWHM of  the stellar image varies
from $2$ to  $3 \ pixel$. The read  out noise and gain of  the CCD are
7.0 $e^-$ and 11.98 $e^-$/ADU, respectively. The fluxes for all of our
programme stars  were extracted by aperture photometry  after the bias
subtraction in  the standard manner  using {\small IRAF}.   Instead of
robust  flat fielding  technique we  used  equation  (3) to  make
uniform response, as mentioned below.

AIMPOL consists  of a half wave  plate and a  Wollaston prism analyzer
placed   in  the  telescope   beam  path   to  produce   ordinary  and
extraordinary  images in  slightly different  directions  separated by
$\sim  27 \  pixels$.  To  re-image the  telescope focal  plane  on the
surface  of  CCD a  focal  reducer  (85mm,  f/1.8) is  placed  between
Wollaston prism and CCD camera with a reduction factor of about 4.7.

By definition,  the ratio R($\alpha$) is given by,
\begin{equation}
R(\alpha) = {I_{e}/I_{\circ} - 1 \over I_{e}/I_{\circ} + 1} = P\ cos(2\theta - 4\alpha)
\end{equation}
which  is  the difference  between  the  intensities  of the  ordinary
($I_{\circ}$) and extraordinary ($I_{e}$) beams to their sum, P is the
fraction of  the total light  in the linearly polarized  condition and
$\theta$  is the  position  angle  of plane  of  polarization.  It  is
denoted by normalized Stokes' parameter {\it q}\ (=Q/I), when the half
wave plate's  fast axis  is aligned to  the reference axis  ($\alpha =
0^\circ$).    Similarly,  the   normalized  Stokes'   parameter  ${\it
  u}\ (=U/I),  {\it q_{1}}\  (=Q_{1}/I), {\it u_{1}}\  (=U_{1}/I)$ are
also  the  ratios  R($\alpha$),  when  the  half  wave  plate  are  at
$22.5^\circ$, $45^\circ$ and  $67.5^\circ$ respectively. In principle,
P and $\theta$ can be  determined by using only two Stokes' parameters
{\it  q} and  {\it u}.   In reality,  the situation  is not  so simple
because of two  reasons (i) the responsivity of the  system to the two
orthogonal  polarization  components may  not  be  same  and (ii)  the
responsivity of the CCD is a  function of position on its surface. So,
the   signals  which  are   actually  measured   in  the   two  images
($I_{\circ}^{\prime}\  and  \  I_{e}^{\prime}$)  may differ  from  the
observed one by the following formula (Ramaprakash 1998)
\begin{equation}
{ I_{e}(\alpha) \over I_{\circ}(\alpha)}={ I_{e}^{\prime}(\alpha) \over I_{\circ}^{\prime}(\alpha)}\times  {F_{\circ} \over F_{e}}
\end{equation}
and the Eq.(1) can be rewrite as

\begin{equation}
R(\alpha) = {(I_{e}^{\prime}/I_{\circ}^{\prime}\times F_{\circ}/{F_{e}})  - 1 \over (I_{e}^{\prime}/I_{\circ}^{\prime}\times F_{\circ}/{F_{e}}) + 1} = P\ cos(2\theta - 4\alpha)
\end{equation}
where $F_{\circ}$ and $F_{e}$ represent the effects mentioned above and the ratio is given by,
\begin{equation}
{ F_{\circ} \over {F_{e}}} = \left[{I_{\circ}^{\prime} ({0}^{\circ}) \over {I_{e}^{\prime} ({45}^{\circ})}}
\times {I_{\circ}^{\prime} ({45}^{\circ}) \over {I_{e}^{\prime} ({0}^{\circ})}}
\times {I_{\circ}^{\prime} ({22.5}^{\circ}) \over {I_{e}^{\prime} ({67.5}^{\circ})}}
\times {I_{\circ}^{\prime} ({67.5}^{\circ}) \over {I_{e}^{\prime} ({22.5}^{\circ})}}\right]^{1/4}
\end{equation}

Substituting the ratio in equation (3) and fitting the cosine curve to
the four values of $R(\alpha)$, the  values of P and $\theta$ could be
obtained.  The  individual errors associated  with the four  values of
$R(\alpha)$  putting as  a weight  while calculating  P,  $\theta$ and
their respective errors.

Four  polarimetric  standard  stars  and four  secondary  polarimetric
standard stars  for null  polarization and for  the zero point  of the
polarization position  angle were taken from Schmidt,  Elston \& Lupie
(1992) and Turnshek \etal (1990) respectively.  The observed degree of
polarization  in  percentage  ($P\%$)  and position  angle  in  degree
($\theta^\circ$)   for  the   polarized  standard   stars   and  their
corresponding values  from Schmidt,  Elston \& Lupie  (1992), Turnshek
\etal (1990) are given in  Table \ref{std_obs}. The observed values of
$P\%$ and  $\theta^\circ$ are  in good agreement  with those  given in
Schmidt, Elston \&  Lupie (1992) and Turnshek \etal  (1990) within the
observational errors.   The observed normalized  stokes parameters $q$
and $u$ in percentage ($q\%,  u\%$) for standard unpolarized stars are
also given in Table  \ref{std_obs}.  The average value of instrumental
polarization is found to be less than $0.05\%$ in all pass-bands.  The
instrumental  polarization  of   AIMPOL  on  the  104-cm  Sampurnanand
Telescope  has been  monitored since  2004 in  different  projects and
found  nearly invariable  in different  pass-bands (Rautela,  Joshi \&
Pandey 2004; Medhi \etal 2007, 2008; Pandey \etal 2009)

AIMPOL have no grid, placed to avoid the overlapping of ordinary image
of  one source  with  the  extraordinary of  an  adjacent one  located
$27~pixels$ away  along the north-south direction. So,  we avoided the
central crowded portion of the cluster. However, the fields are chosen
in such a  manner to include maximum number of  member stars.  We also
had a large number of sources which are not members but are present in
the  fields  observed.  All  the  sources  were  manually checked  and
rejected in case of an overlapping image.

\begin{table}
\centering
\caption{Results of previous polarization measurements carried out 
by Serkowski (1965) in the direction of NGC 6823.}
\begin{minipage}{500mm}
\label{lit_res}
\begin{tabular}{llllll}
\hline
\hline
ID(B) &$P_{B}\pm\epsilon(\%)$&$\theta_B$&$P_{V}\pm\epsilon(\%)$&$\theta_V$&$M_{P}$\\
(1)&(2)&(3)&(4)&(5)&(6)\\
\hline
\multicolumn{6}{|c|}{Serkowski (1965)}\\
\hline

  3  &   4.14 $\pm$ 0.23 &  06   &  4.19 $\pm$ 0.23 &   06 &  0.97  \\
  9  &   4.51 $\pm$ 0.32 &  172  &  4.84 $\pm$ 0.32 &   175&  0.97  \\
 11  &   5.11 $\pm$ 0.46 &  05   &  4.47 $\pm$ 0.46 &   04 &  0.97  \\
 13  &   2.95 $\pm$ 0.14 &  11   &  3.32 $\pm$ 0.14 &   09 &  0.95  \\
 17  &   2.81 $\pm$ 0.51 &  13   &  3.36 $\pm$ 0.51 &   16 &  0.97  \\
 54  &  \ ---          &\ ---  &  0.83 $\pm$ 0.28 &   03 &\ ---   \\  
 68  &   4.38 $\pm$ 0.46 &  14   &  5.34 $\pm$ 0.46 &   12 &  0.96  \\
 73  &  \ ---          &\ ---  &  0.23 $\pm$ 0.18 &   30 &\ ---   \\
 74  &   2.58 $\pm$ 0.46 &  13   &  3.64 $\pm$ 0.46 &   12 &  0.97  \\
 78  &   3.41 $\pm$ 0.55 &  02   &  4.10 $\pm$ 0.55 &   09 &  0.93  \\
189  & \ ---           &\ ---  &  0.23 $\pm$ 0.14 &   52 & \ ---  \\
$BD+23^{\circ}3745$   &  6.40 $\pm$ 0.14 &  27  & 6.63 $\pm$ 0.14 &  24 & 0.88  \\
\hline
\end{tabular}
\end{minipage}
\end{table}     

\section{Results and Discussion} 

The  results of  our polarimetric  observations towards  NGC  6823 are
presented  in the  Table  \ref{tab_2} and  Table \ref{tab_3}.   Star's
identification numbers  ID(M) and  ID(B) are given  in column 1  and 2
following  this observation and  Barkhatova (1957)  respectively.  The
instrumental  magnitudes obtained in  $V$ filter  are given  in column
3. The  measured values of polarization $P\%$  and their corresponding
error  $\epsilon \%$  in $B,  V  $ and  $R_{c}$ filters  are given  in
columns 4, 6 and 8, respectively.  The polarization position angle (of
the $E$  vector) $\theta^\circ$ and the  corresponding error $\epsilon
^\circ$ in $B, V$ and $R_{c}$ filters are given in columns 5, 7 and 9,
respectively.  The position angles in the equatorial coordinate system
are measured  from the  north increasing eastward.  Columns 10  and 11
represent  the  membership  probabilities  $M_{P}(E)$  and  $M_{P}(D)$
according    to    Erickson   (1971)    and    Dias   \etal    (2006),
respectively. Stars  with $M_{P}\geq  0.50$ are considered  as cluster
members in this study.

The previous  polarimetric measurements of  stars in the  direction of
NGC  6823  carried out  by  Serkowski  (1965)  is presented  in  Table
\ref{lit_res}.  Serkowski (1965) observed 22 stars in the direction of
NGC 6823 using $B$ and $V$ photometric bands. Out of 22 stars observed
by Serkowski (1965) we have included  only 12 stars in our study which
have  the  available  R.A.  and   Dec.   In  column  1,  we  give  the
identification numbers which are adopted from Barkhatova (1957). $P\%$
and $\theta^\circ$ in B and V filters are given in columns 2, 3, 4 and
5, respectively. We  converted $p$ to $P$ per  cent using the relation
$P$  per  cent$=46.05p$ (Whittet  1992).   In  column  6 we  give  the
membership probabilities of stars obtained from Erickson (1971).

Figure  \ref{NGC6823_V_dss.ps}  presents  the  sky projection  of  the
V-band polarization  vectors for the 104  stars observed by  us in NGC
6823 (R band  image is reproduced from Digitized  Sky Survey II).  The
polarization vectors  are drawn at  the center of the  observed stars.
The  length  of  the   polarization  vector  is  proportional  to  the
percentage  of polarization  in V  band ($P_{V}$)  and it  is oriented
parallel  to  the  direction  of corresponding  observed  polarization
position angle  in degree  in V band  ($\theta_{V}$). The  dashed line
represents  the  Galactic parallel  at  $b=-0.14^{\circ}$ inclined  at
$\sim32  ^{\circ}$  with  respect   to  the  north.   The  stars  with
$M_{P}\geq  0.50$ are identified  using closed  star symbols  in white
colour. Polarization vectors for 12 stars observed by Serkowski (1965)
are shown in white colour.

In Figure \ref{NGC6823_B_dss.ps}, we present the sky projection of the
B-band polarization vectors  for the 104 stars observed  by us towards
NGC  6823  (vectors in  black  colour)  along  with the  results  from
Serkowski   (1965)   (vectors  in   white  colour)   and  in   Figure
\ref{NGC6823_R_dss.ps}  we  present only  the  sky  projection of  the
$R_{c}$-band polarization vectors for the  104 stars observed by us in
NGC 6823 (vectors in black colour), because there is no observations by
Serkowski (1965) in R filter.

There  are eight  stars in  common  between the  Serkowski (1965)  and
present  observations. The  polarization and  position angles  for the
eight common  stars in B and  V filters seems to  be consistent within
the uncertainty in both the  observations.  From the sky projection of
the polarization vectors for 104 stars observed by us in all the three
filters,  it is  clearly exist  that the  polarization vectors  of the
stars distributed about the Galactic  plane and less than the Galactic
plane (especially those located at the center of the cluster).

Figure \ref{dif_pa.ps} shows the  distribution of the polarization and
position angle in $V$ filter  ($P_V$ and $\theta_V$) for the 42 member
stars  with the  radial distance  from  the cluster  center.  We  have
chosen  $(R.A.(J2000):  19^{h}  43^{m} 09^{s},  Dec(J2000):+23^{\circ}
18^{m}  00^{s}$ as  the center  of the  cluster taken  from Kharchenko
\etal  2005,  where they  use  the  approximation  of maximum  surface
density of cluster members for locating the cluster centre.  The upper
and lower left panel of the Fig. \ref{dif_pa.ps} show the distribution
of  position angle  and polarization  in  $V$ filter  with the  radial
distance from the center of the cluster. The position angles are found
to show  a systematic  change while the  polarization found  to reduce
towards  the outer  parts of  the cluster.   In both  the plots  it is
noticable  that  the member  stars  show  preferentially two  separate
distributions which  are shown  by the histograms  in upper  and lower
right panel  of the Fig.  \ref{dif_pa.ps}. From the cluster  center to
radial distance  of 4 arcmin  the stars are  showing a trend  of lower
position angle and higher  polarization, black filled star symbol than
the stars  lying at  radial distance of  above 4 arcmin,  black filled
circle    in     the    upper     and    lower    left     panel    of
Fig. \ref{dif_pa.ps}.  Stone (1988) also found a  boundary at $r\simeq
3.5$ arcmin and he defines  the region from cluster center to $r\simeq
3.5$  arcmin as  the nucleus  and above  $r\simeq 3.5$  arcmin  as the
coronal region of the cluster NGC 6823.  Other authors like Barkhatova
(1957) found the radius for the nuclear region as $r\simeq 4.0$ arcmin
while Turner  (1979) quote  it as $r\simeq  2.5$ arcmin.   In general,
every  cluster  consists  of  two   main  regions,  a  nucleus  and  a
corona. Nucleus is the densest,  central part of the cluster, which is
perceived  directly by  our  eye as  a  cluster. Corona  is an  outer,
extended,  less dense  region around  the cluster.   For  many distant
clusters, the coronas are lost against the rich star field.

The right upper and lower panel of the Fig.\ref{dif_pa.ps} present the
distribution of 42 member stars  belong to the nucleus and the coronal
region  shown by the  dashed and  the filled  line histogram.   In NGC
6823, out  of 42 member  stars 24 stars  belong to the nucleus  and 18
stars  belong  to  the  coronal region.  Two  significantly  different
distributions of the polarization  and the position angle are followed
by the  stars belong  to nucleus  and coronal region  as shown  in the
histograms  in  Fig.\ref{dif_pa.ps}.    The  weighted  mean  value  of
polarization ($P_V$) and position angle ($\theta_V$) in $V$ filter are
3.66\pmi0.02\%,  8$^\circ$\pmi1 for  the nucleus,  and 3.10\pmi0.02\%,
25$^\circ$\pmi1  for  the coronal  region  respectively.  The  average
value of  the position angle of  the coronal region is  more closer to
the inclination  of the Galactic parallel  ($\sim32^{\circ}$) than the
nuclear region.   The distribution  of polarization vectors  about the
Galactic plane indicates  that the dust grains present  in the coronal
region are mostly aligned by a magnetic field which is nearly parallel
to the  direction of  the Galactic Disk.   Whether in the  nucleus the
distribution  of polarization  vectors  less than  the Galactic  plane
indicates that a second component  of magnetic field which is slightly
less inclined to the Galactic Disk could also be present.

In  the radial  distribution plots  (Fig.\ref{dif_pa.ps}), it  is also
noticed that  the polarization ($P_V$) data points  are more scattered
in the  nucleus as compared to  the coronal region  while the position
angle ($\theta_V$) data points  are showing the same behaviour (highly
scattered) in both regions. The  highly scattered $P_V$ data points in
the nuclear region indicate that  the density and distributions of the
intracluster  dust/materials  may  be  higher  and  more  differential
respectively  than  the coronal  region.   The  presence of  different
generation of dust particles and different component of local magnetic
field may be the cause  for highly scattered $\theta_V$ data points in
the radial plot for both regions.

\begin{figure}
\resizebox{7.5cm}{7.0cm}{\includegraphics{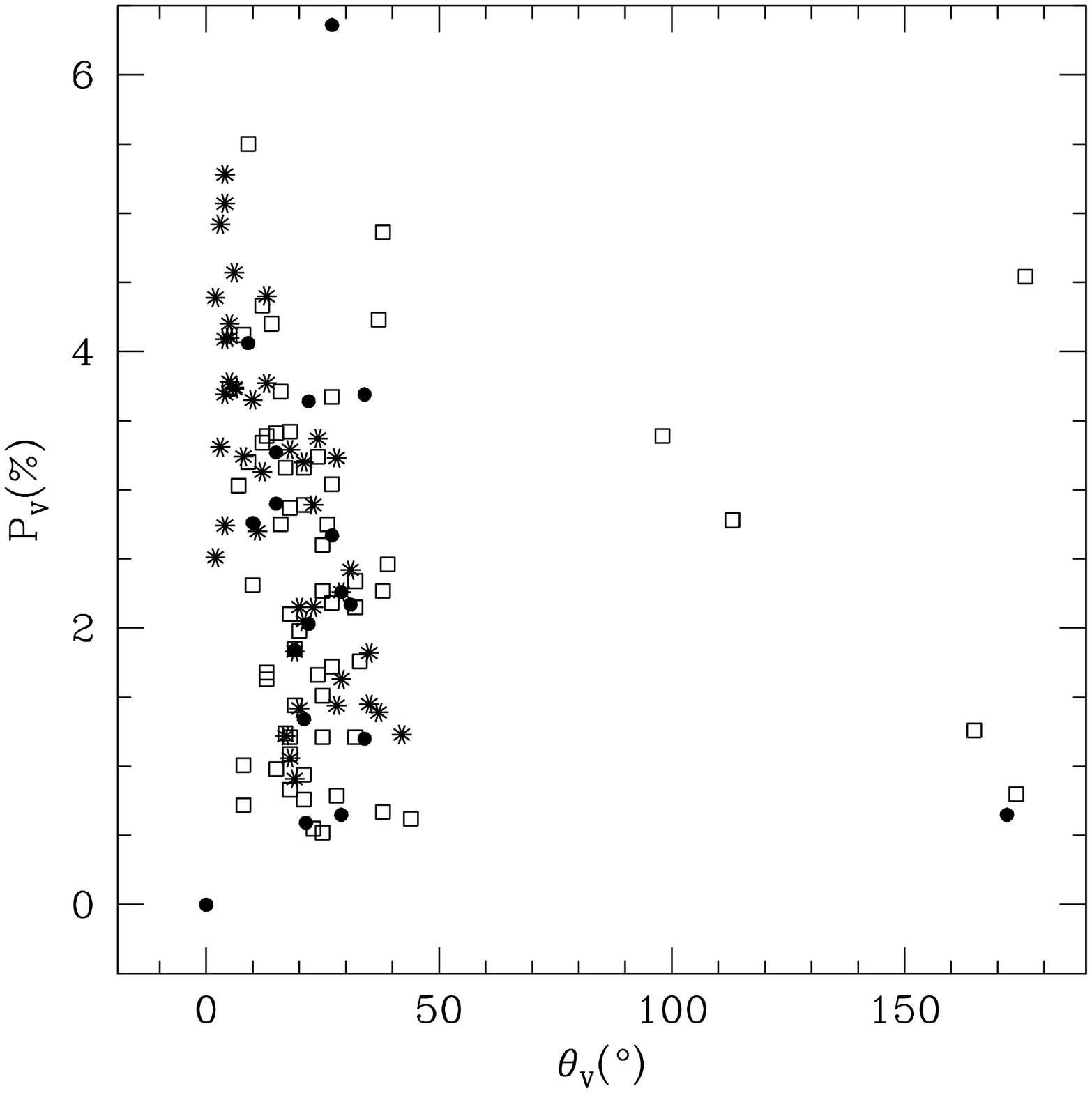}}
\caption{The $P_{V}$ vs. $\theta_{V}$  plot for 104
stars observed by us in the direction of NGC 6823 are shown using open
squares. Stars taken from  Heiles (2000) are  shown by  using  filled  black  circles. 
The stars  with $M_p  \geq 0.50$  are identified  using 
star symbols.}
\label{p_pa.ps}
\end{figure}

To  check  the  consistency  of  our results,  we  use  Heiles  (2000)
catalogue  which   has  a   compilation  of  over   9000  polarization
measurements. We found 19  stars with polarization measurements in $V$
band, within a circular  region of radius $\simeq2^{\circ}$ around NGC
6823.    Figure  \ref{p_pa.ps}   presents  $P_{V}$   vs.  $\theta_{V}$
plot. Our  results are represented by open  squares. The polarizations
and  position angles  in visual  filter taken  from Heiles  (2000) are
represented by  filled black  circles. The stars  with $M_{P}\geq0.50$
are identified using star symbols.   The stars from Heiles (2000) show
degree  of  polarization ($P_H$)  in  the  ranges  from $\sim0.01$  to
$\sim6.36$\%.  The mean value of $P_H$ and position angle ($\theta_H$)
are       $\simeq2.21\pm0.86\%$      and      $\simeq28\pm10^{\circ}$,
respectively. In our  observation, the $P_{V}$ of the  stars are found
to be in the ranges from  $\sim0.52$ to $\sim5.50\%$. The stars make a
clustering towards the lower position angle. The mean value of $P_{V}$
and   $\theta_{V}$   are   found   to   be   $\sim2.58\pm0.39\%$   and
$\sim26\pm4^{\circ}$, respectively.

The  mean values  of polarization  and position  angles for  a smaller
region  of $\sim25^{\prime}\times25^{\prime}$  centering at  NGC 6823,
obtained by  us are nearly similar  to those for a  larger region from
Heiles (2000).   The cause of polarization is,  therefore, must likely
be  due to the  contribution from  the dust  grains distributed  in an
extended  structure  closer to  us  and  from  the small  intracluster
contribution.  Consequently,  the knowledge about  the distribution of
interstellar dust towards the direction of NGC 6823 is very important,
to interpret the our polarimetric results.

\subsection{Distribution of interstellar matter in the region of NGC 6823}

The color excess E(B-V) of this cluster for main sequence stars varies
from  $0^{m}.60 $  to $1^{m}.16$  (Sagar \&  Joshi 1981,  Sagar 1987).
Moreover, a  relatively larger  value of color  excess is found  at the
center and  along the  diagonal joining the  North-West corner  to the
South-East corner of the cluster  (Sagar \& Joshi 1981).  In NGC 6823,
there  is a  slight tendency  that  the E(B-V)  for the  stars in  the
eastern  part of  the cluster  exhibit a  large value  of  $R_V$.  The
cluster NGC  6823 is surrounded by  a reflection nebula  NGC 6820. The
extinction is highest at the eastern part of the cluster as it is
the direction of the reflection nebula.  But excluding super-giant the
reddening law for the whole cluster can be characterized by $R_V = 3.2
\pm 0.1$ (Guetter 1992).

\begin{figure}
\resizebox{8.5cm}{7.5cm}{\includegraphics{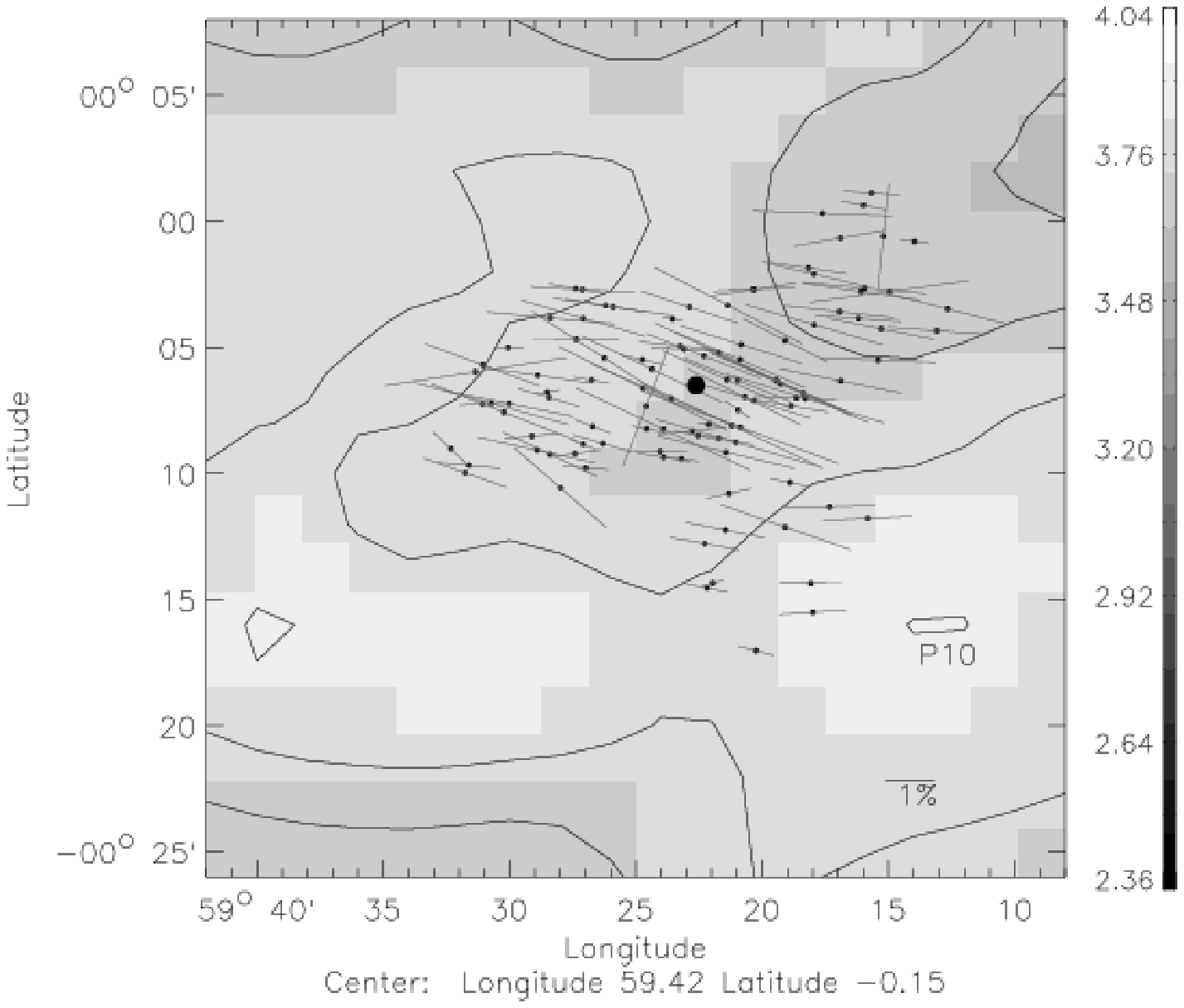}}
\caption{The   high   resolution   extinction   map  of   the   region
  ($30^{\prime}\times30^{\prime}$) produced by  Dobashi et al.  (2005)
  using  the optical database  `Digitized Sky  Survey I'  and applying
  traditional star count technique. We overlay V band results from our
  observations  using  vectors  drawn  in  black  colours.  The  clump
  identified by  Dobashi et al.   (2005) in this region  is identified
  and labeled  as P10.  The center of  the cluster ($l=59.40^{\circ}$,
  $b=-0.14^{\circ}$) is identified using black circle.}
\label{dobashi_v}
\end{figure}

By using  the optical database  `Digitized Sky Survey I'  and applying
traditional  star  count technique,  Dobashi  et  al. (2005) 
produced extinction  maps of  the entire region  of the Galaxy  in the
galactic  latitude range  $|b|\leq40^{\circ}$. We  have used  their fits
images  of  the  extinction  map  of the  field  containing  NGC  6823
available on-line
\footnote{http://darkclouds.u-gakugei.ac.jp/astronomer/astronomer.html}.

The high resolution  extinction map overlaid with V  band results from
our observations  is shown in Figure  \ref{dobashi_v}.  We transformed
all position angles  measured relative to the equatorial  north to the
galactic  north using  the relation  given by  Corradi et  al. (1998),
because the $A_{V}$ maps are in galactic coordinates. The black
circle  identifies  the  center  of  the  cluster  ($l=59.40^{\circ}$,
$b=-0.144^{\circ}$).  The black and  white colour-bar on the right hand
side of  the extinction map shows  the range of $A_{V}$  values in the
figure.  The contours are plotted at $A_{V}$=0.5 to 4 with an interval
of 0.5 magnitude.
\begin{figure}
\resizebox{8.5cm}{6.5cm}{\includegraphics{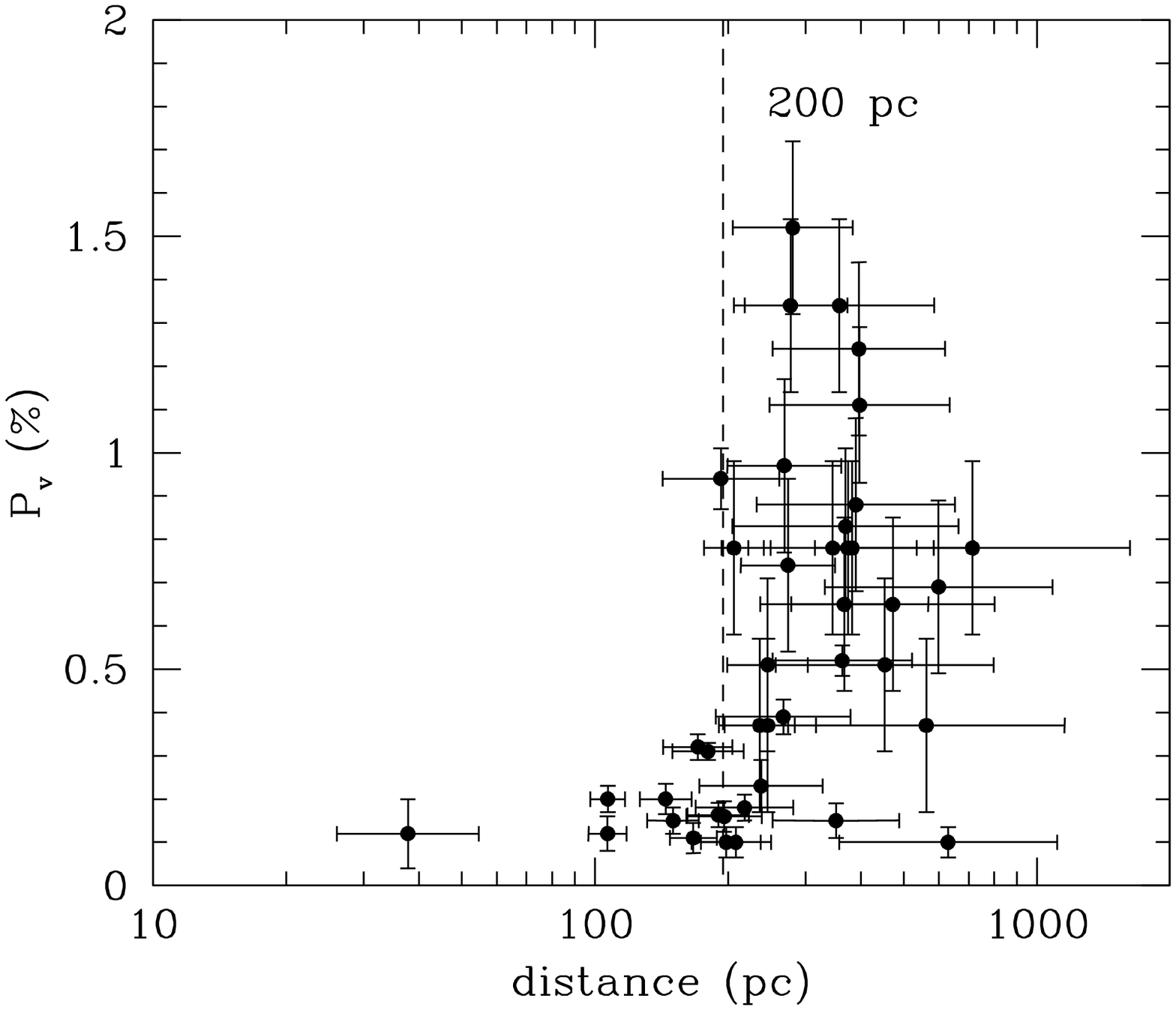}}
\caption{Distance versus $P_V\%$  plot, data obtained from Heiles (2000) and 
Perryman \etal (1997). The dotted line is drawn at
200 pc to show a sharp increase in $P_V\%$ value at these
distance.}
\label{p_dis.ps}
\end{figure}

\begin{figure*}
\resizebox{12cm}{12cm}{\includegraphics{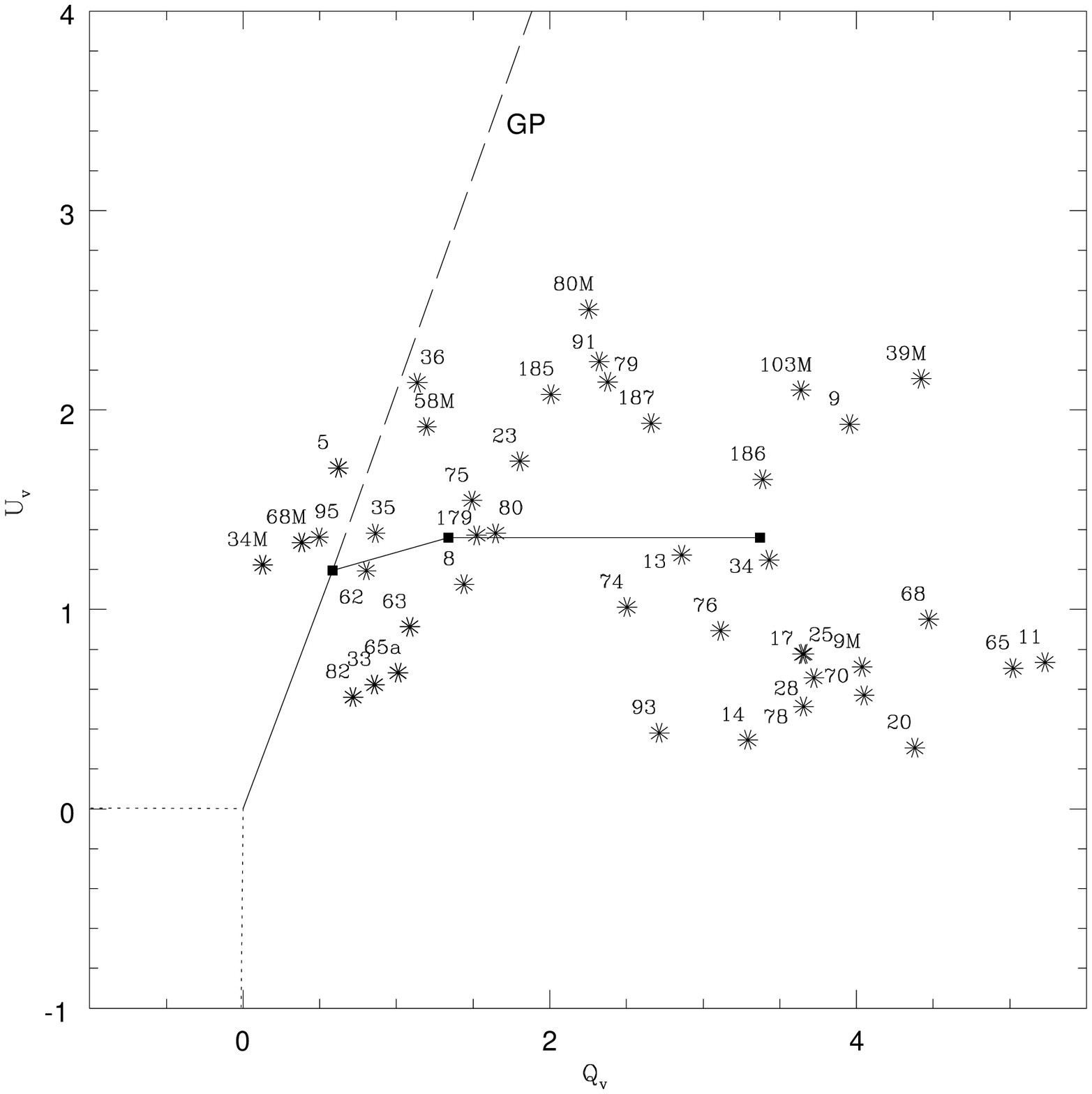}}
\caption{The $U_{V}$ vs. $Q_{V}$  plot for 42 member stars
observed by us in the direction of NGC 6823. The dashed line is in the direction of
the Galactic Plane (GP). The solid line is the interpretation of the evolution
of polarization through the dust layer between the cluster and the Sun.}
\label{UQ.ps}
\end{figure*}

The  extinction towards  the location  of the  cluster  (black circle)
shows relatively low ($A_{V}\sim3$)  values.  But the outer regions of
the cluster especially towards the  south and the east, the extinction
increases up to $\sim4$ magnitude.  One clump identified by Dobashi et
al.(2005)   in   this   region   is   labeled   as   P10   in   Figure
\ref{dobashi_v}. There  is only one  dark cloud LDN 791  identified by
Lynds (1962) located radially $40^{\prime}$ away from the clump P10 in
south east  direction.  The value of extinction  estimated by previous
observers (Guetter 1992, Feinstein  1994, Sagar \& Joshi 1981) towards
the center of the cluster is nearly agree with Dobashi et al.(2005).

The Columbia 1.2-m millimeter-wave  telescope CO emission survey in NGC
6823  by Leisawitz, Bash  \& Thaddeus  (1989) identify  five molecular
clouds around it at the  cluster distance.  The velocity and masses of
the clouds ranges from $23.7\ km s^{-1}$ to $34.8 \ km s^{-1}$ and $11
\times 10^3 M_\odot$ to  $110 \times 10^3 M_\odot$, respectively.  The
Balloon-borne  Large Aperture  Sub-millimeter  Telescope (BLAST)  also
detect sixty compact sub-millimeter sources simultaneously at 250, 350
and 500 $\mu m$ (Chapin \etal  2008) towards NGC 6823.  Of these sixty
compact sub-millimeter  sources, forty  nine of them  are found  to be
associated with NGC 6823.

To  have  an   idea  about  the  distances  to   the  foreground  dust
concentration,  it is  very  important to  study  the distribution  of
polarization at different distances  in that particular line of sight.
So,  to investigate the  polarization properties/distribution  of dust
grains towards  NGC 6823 located  at different distances,  we selected
all the stars from Heiles (2000) within a circle of radius $5^{\circ}$
around  NGC  6823  with  available polarization  data  ($P_{H}$).  The
parallax  measurements for  the stars  are obtained  from  the \textit
{Hipparcos and Tycho Catalogs} (Perryman \etal 1997 and H$\phi$g \etal
1997).  Applying 2$\sigma$ rejection  limit criterion to the parallax,
we found  parallax measurements available only for  45 stars, covering
maximum distances up to 800  pc.  Stars which showed peculiar features
and emissions in their spectrum, as given by SIMBAD, are rejected.

Figure  \ref{p_dis.ps} presents the  degree of  polarization ($P_{H}$)
versus distance plot. The stars located closer to us show low value of
$P_{H}$ ($\lesssim0.2\%$)  than the stars located  beyond $\sim200$ pc
which show relatively high values of $P_{H}$ in the range from $0.2\%$
to $1.5\%$ with  a sharp jump in polarization  occurring at $\sim 200$
pc.  Only one  star is available in between  $\sim10$ to $\sim100$ pc,
so from the Figure \ref{p_dis.ps} we can infer only a maximum distance
of $200$ pc to the dust grains responsible for the observed sharp jump
and assign a  maximum distance of $200$ pc to the  first layer of dust
towards NGC 6823.

Since, most of the  extinction/reddening is produced by the foreground
interstellar  material so, it  is necessary  to investigate  about the
evolution  of  the  interstellar  environments  from the  Sun  to  the
cluster. In  Figure \ref{UQ.ps}, we have plotted  the Stokes parameter
$Q_V$  ($=P_Vcos 2\theta_V$)  and $U_V$  ($=P_Vsin 2\theta_V$)  in $V$
filter,  for each  of the  42 observed  member stars.  In  figure, the
coordinates   $Q_V$=0  and  $U_V$=0   represent  the   dustless  solar
neighbourhood. The points lying at  other parts in the figure indicate
the direction  of the polarization vector  as seen from  the sun.  The
member stars  over this region roughly segregate  into three different
polarimetric group  according to  distribution in Stokes  plane. Solid
lines in  Figure \ref{UQ.ps} represent  the changing direction  of the
vector $P_V$  while connecting the  weighted mean values of  $Q_V$ and
$U_V$  for  the possible  three  different  groups.  The nearby  group
consists of  ten member  stars (namely, \#5,  \#33, \#35,  \#62, \#63,
\#65a,  \#82, \#95,  \#68M, \#34M;  Identification number  suffix  by
``M'' is  according to  ID(M) and others  according to ID(B)),  with a
weighted mean  value of polarization  $1.34 \pm 0.04 \%$  and position
angle $33 \pm 1^\circ$. This  nearby group lies almost parallel to the
Galactic  Plane. The  next group  has six  member stars  (namely, \#8,
\#36,  \#75,   \#80,  \#179,  \#58M)  with  weighted   mean  value  of
polarization and position angle $1.93  \pm .05 \%$ and $23 \pm 1^\circ
$, respectively.  The remaining  twenty six member stars (namely, \#9,
\#11,  \#13, \#14,  \#17, \#20,  \#23, \#25,  \#28, \#34,  \#65, \#68,
\#70, \#74, \#76,  \#78, \#79, \#91, \#93, \#185,  \#186, \#187, \#9M,
\#39M, \#80M,  \#103M) are  in the third  group, which has  a weighted
mean value of  polarization $3.64 \pm 0.01 \%$  and position angle $11
\pm 1^\circ$.

The evolution of polarization towards NGC  6823 is likely to be due to
patchy distribution  of dust and  the core may  be behind a  low dense
layer of dust or a hole.  Some of the authors who made the photometric
studies of NGC 6823 believe that towards this cluster the distribution
of  dust is patchy,  \eg; Guetter(1992)  indicates that  the reddening
towards this  cluster may  be constant in  small spatial areas  but is
highly variable across the face  of the cluster and ranges from $1.07$
to $0.64$ mag.   This absorption variability has been  also noticed by
many earlier  investigators and  can easily be  seen by  examining the
color-magnitude  and  color-color plots  as  published  by Hoag  \etal
(1961).

From the optical photometric study, Neckel and Klare (Neckel \& Klare,
1980) found  that more than half  of the total  extinction towards the
cluster  NGC 6823  comes  from the  matter  lying close  to  us, at  a
distance  between  0.2   to  0.5  kpc.   The  presence   of  a  nearby
interstellar matter at a distance of $\sim 300$ pc in the direction of
cluster was also recently confirmed by Fresneau \& Moiner (1999). They
noticed a systematic  absorption of about $1^{m}.5$ in  V band in this
region. The absorbing matter in this region may be probably located at
a  depth of  the Vulpecula  rift  molecular cloud  (Dame \&  Thaddeus,
1985).  Therefore,  we believe that the  observed polarization towards
NGC 6823  is mainly  due to  layers of nearby  aligned dust  grains of
patchy structure associated with the above mentioned clouds.

\begin{figure}
\resizebox{8cm}{8cm}{\includegraphics{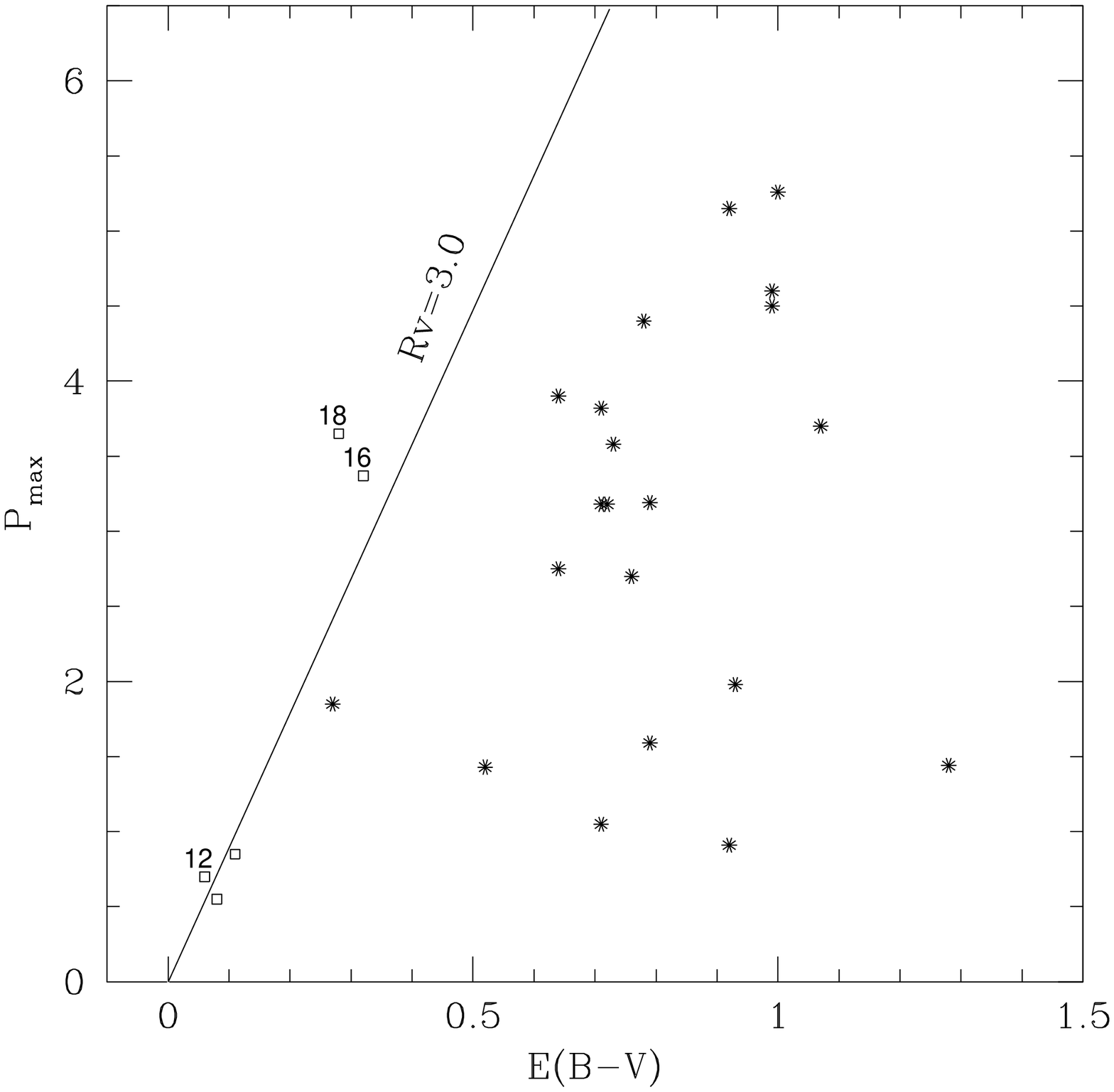}}
\caption{Polarization  efficiency diagram. Using  Rv=3.0, the  line of
maximum efficiency drawn.   The stars observed by us  in the direction
of NGC 6823 are  shown using open black square. The stars  with $M_p  \geq 0.50$ 
are identified  using star symbol.}
\label{eff}
\end{figure}
\begin{figure}
\resizebox{8cm}{8cm}{\includegraphics{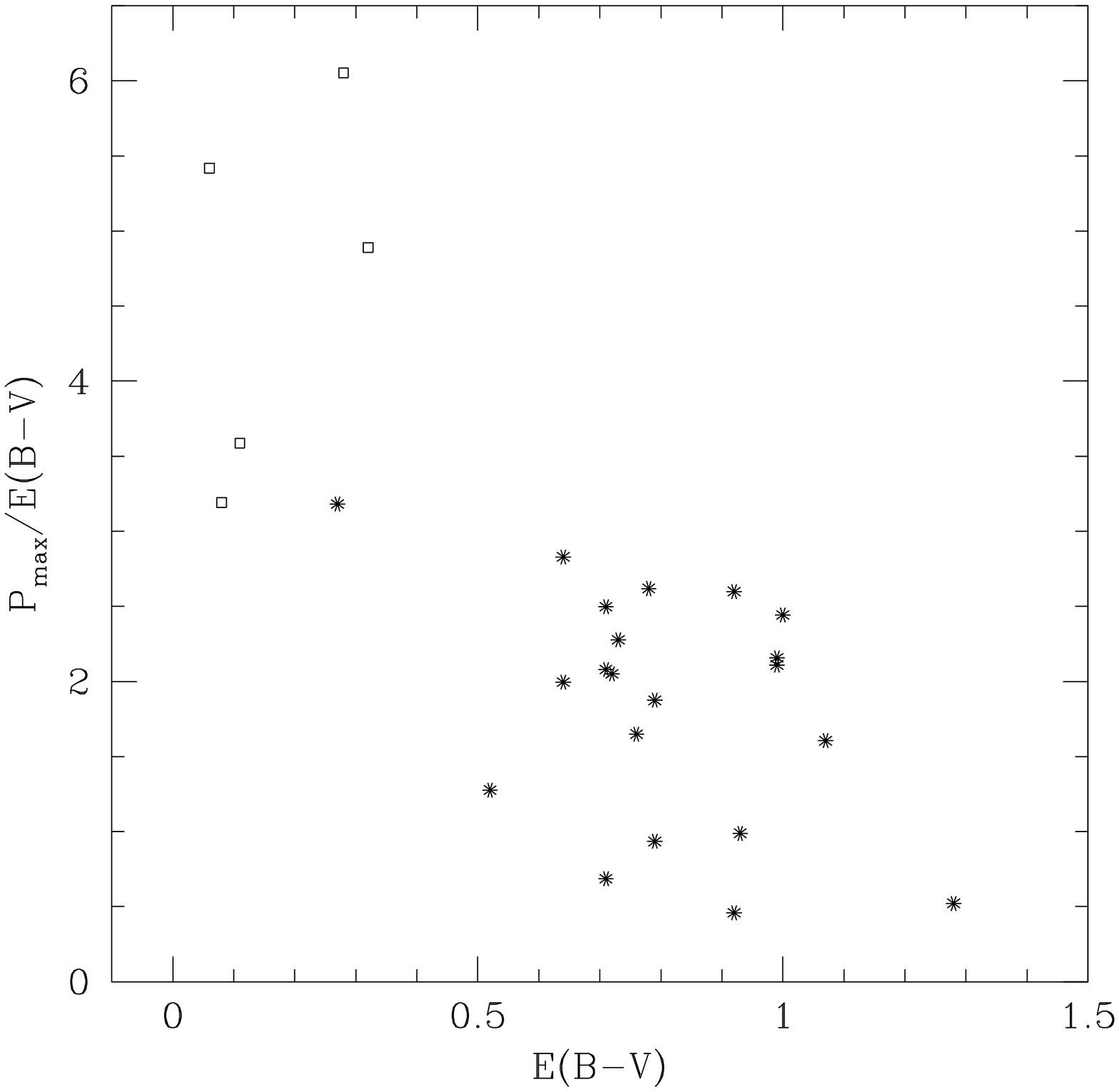}}
\caption{$P_{max}/E(B-V)$ plotted as a function of $E(B-V)$. The stars
observed  by us  in the  direction of  NGC 6823  are shown  using open black
square. The  stars with $M_p  \geq 0.50$ are
identified using star symbol. }
\label{eff13}
\end{figure}

\subsection{Serkowski Law}

The maximum wavelength  ($\lambda_{max}$) and the maximum polarization
($P_{max}$)  both   are  the  functions  of   optical  properties  and
characteristic  of  particle size  distribution  of  the aligned  dust
grains (McMillan  1978, Wilking  et al. 1980).   Moreover, it  is also
related to  the interstellar  extinction law (Serkowski,  Mathewson \&
Ford 1975, Whittet \& van Breda 1978, Coyne \& Magalhaes 1979, Clayton
\&  Cardelli  1988).   The  $\lambda_{max}$ and  $P_{max}$  have  been
calculated  by fitting  the  observed  polarization in  the  B, V  and
$R_{c}$ band-passes to the standard Serkowski's polarization law;

\begin{equation}
P_{\lambda}/    P_{max}   =\    exp    \left[-\   k    \   ln^{2}    \
(\lambda_{max}/\lambda) \right]
\end{equation}
and adopting the  parameter $k = 1.15 $ (Serkowski  1973). In the fits
the degree of freedom is adopted  as one.  Though there are only three
data points, the wavelength covered ranges  from 0.44 to 0.66 $\mu m $
and all the $\lambda_{max}$ found to fall within this range. Since, we
have  enough  wavelength coverage,  the  fit  is  reasonably fine  but
sometimes it over estimates the value of $\sigma_1$.  For each star we
computed $\sigma_1$ parameter (the unit  weight error of the fit).  If
the polarization  is well represented by  the Serkowski's interstellar
polarization law, $\sigma_1$ should not  be higher than 1.6 due to the
weighting scheme.  A higher value  could be indicative of the presence
of intrinsic  polarization.  The $\lambda_{max}$ can also  give us the
clue  about  the  origin   of  polarization.   The  stars  which  have
$\lambda_{max}$  lower  than the  average  value  of the  interstellar
medium  ($0.55 \pm  0.04 \  \ \mu  m$, Serkowski  \etal 1975)  are the
probable  candidates to  have an  intrinsic component  of polarization
(Orsatti, Vega \& Marraco 1998).  The dispersion of the position angle
($\overline{\epsilon}$) for each star  normalized by the mean value of
the position angle errors is  the another tool to detect the intrinsic
polarization.   The values  obtained  for $P_{max}$,  $\lambda_{max}$,
$\sigma_1$  and  $\overline{\epsilon}$  together  with  ID(M),  ID(B),
R.A.(2000J) and Dec(2000J)  for all the 104 observed  stars with their
respective errors are given in Table 5 and 6.  Column 9 in the Table 5
and 6  represent $E(B-V)$,  available only for  21 member stars  and 5
non-member stars. In  that table the values of  $E(B-V)$ sufix by (G),
(S)  and (E) taken  from Guetter  1992, Stone  1988 and  Erickson 1971
respectively.

Out of 104 stars observed by us, only for one non-member star (namely,
\#57M) and three member stars  (namely, \#08M, \#40M and \#102M) found
the value of $\sigma_1$ above the  limit of 1.6. The dispersion in the
position  angle $\overline{\epsilon}$ is  found to  be higher  for one
member star (namely, \#25M) and eight non-member stars (namely, \#15M,
\#21M,   \#45M,  \#70M,   \#78M,  \#87M,   \#88M  and   \#101M).   The
$\lambda_{max}$ for  above mentioned thirteen stars  detected by using
two criteria $\sigma_1$ and  $\overline{\epsilon}$ are nearly equal to
the interstellar  medium except  the non-member star  \#78M.  Although
the  non-member   star  \#78M  is   found  to  have  lower   value  of
$\lambda_{max}$  and  higher value  of  $\overline{\epsilon}$ but  the
$\sigma_1$ is  lower than the threshold.   Therefore, the polarization
towards  NGC6823 is  found  to  be mainly  because  of the  foreground
interstellar dust grains.

If the polarization is mainly  dominated by the dust particles present
in the diffuse interstellar medium the value of $\lambda_{max}$ should
be  nearly equal  to 0.55  $\pm$ 0.04  $\mu$m.  The  weighted  mean of
$\lambda_{max}$  for the observed  member and  nonmember stars  of NGC
6823 are obtained as $ 0.53\pm 0.01  \ \mu m$ and $ 0.54\pm 0.01 \ \mu
m$, respectively.   The nearly  similar values of  $\lambda_{max}$ for
the member  and the  non-member stars imply  that the lights  from the
both member  and non-member stars encountering the  same population of
foreground  dust  grains.  Therefore,  the characteristic  grain  size
distribution as  indicated by the  polarization study of stars  in NGC
6823 is nearly same as  that for the general interstellar medium.  The
weighted  mean  of  the   maximum  polarization  for  the  member  and
non-member  stars are  found as  $ 2.82\pm  0.01 \  \%$ and  $ 1.76\pm
0.01\ \%$, respectively.

\begin{table*}
\centering
\begin{minipage}{310mm}
\caption{The $P_{max}$, $\lambda_{max}$, $\sigma_1$ \& $\overline{\epsilon}$ for the observed data in NGC 6823.}
\begin{tabular}{llllllllll}
\hline
\hline
ID(M)$^\dag$& ID(B)$^\dag$$^\dag$& R.A(2000J) & DEC(2000J)&{$P_{max}$\pmi $\epsilon$} $(\%)$ &  $\lambda_{max}$\pmi$\epsilon$ ($\mu m$)&$\sigma_1$  & $\overline{\epsilon}$& $E(B-V)$ \\
(1)&(2)& (3)&(4)&(5)&(6)&(7)&(8)&(9)\\
\hline
01  & --- &  19\ 42\ 51.00 & 23\ 18\ 42.2  &  2.08 \pmi 0.11 &  0.50 \pmi 0.05  &  0.36 & 0.48  &  ---  \\ 
02  & 20  &  19\ 42\ 54.82 & 23\ 18\ 59.6  &  4.40 \pmi 0.01 &  0.56 \pmi 0.01  &  0.18 & 0.44  &  0.78(G) \\ 
03  & 74  &  19\ 42\ 58.40 & 23\ 20\ 16.9  &  2.70 \pmi 0.01 &  0.53 \pmi 0.01  &  0.75 & 0.44  &  0.76(G) \\ 
04  & 73  &  19\ 43\ 01.60 & 23\ 20\ 37.7  &  0.53 \pmi 0.01 &  0.55 \pmi 0.01  &  0.10 & 0.21  &  ---  \\ 
05  & 18  &  19\ 42\ 59.58 & 23\ 17\ 45.9  &  3.65 \pmi 0.08 &  0.55 \pmi 0.03  &  0.59 & 0.81  &  0.28(G) \\ 
06  & 23  &  19\ 42\ 55.14 & 23\ 16\ 20.4  &  2.52 \pmi 0.01 &  0.58 \pmi 0.01  &  0.02 & 0.60  &  ---  \\ 
07  & 17  &  19\ 43\ 01.94 & 23\ 17\ 30.3  &  3.90 \pmi 0.01 &  0.54 \pmi 0.01  &  1.09 & 0.33  &  0.64(S) \\ 
08  & 78  &  19\ 43\ 02.76 & 23\ 18\ 22.8  &  3.58 \pmi 0.14 &  0.60 \pmi 0.04  &  2.68 & 0.44  &  0.73(G) \\ 
09  & --- &  19\ 43\ 02.06 & 23\ 15\ 37.4  &  4.32 \pmi 0.08 &  0.53 \pmi 0.02  &  0.75 & 0.33  &  ---  \\ 
10  & 27  &  19\ 43\ 02.84 & 23\ 14\ 49.2  &  2.16 \pmi 0.02 &  0.58 \pmi 0.01  &  0.22 & 0.86  &  ---  \\ 
11  & 26  &  19\ 43\ 04.45 & 23\ 14\ 49.5  &  2.02 \pmi 0.02 &  0.51 \pmi 0.01  &  0.58 & 0.67  &  ---  \\ 
12  & 13  &  19\ 43\ 06.85 & 23\ 16\ 12.6  &  3.18 \pmi 0.03 &  0.54 \pmi 0.02  &  0.99 & 0.89  &  0.71(G) \\ 
13  & 28  &  19\ 43\ 01.53 & 23\ 14\ 40.8  &  3.83 \pmi 0.01 &  0.60 \pmi 0.01  &  0.02 & 0.53  &  ---  \\ 
14  & 75  &  19\ 43\ 05.01 & 23\ 19\ 50.0  &  2.25 \pmi 0.04 &  0.56 \pmi 0.03  &  0.65 & 0.19  &  ---  \\ 
15  & 77  &  19\ 43\ 05.17 & 23\ 19\ 37.9  &  1.72 \pmi 0.01 &  0.54 \pmi 0.01  &  0.03 & 1.20  &  ---  \\ 
16  & 76  &  19\ 43\ 04.49 & 23\ 18\ 49.4  &  3.28 \pmi 0.02 &  0.56 \pmi 0.01  &  0.69 & 0.67  &  ---  \\ 
17  & 16  &  19\ 43\ 06.03 & 23\ 17\ 33.1  &  3.37 \pmi 0.20 &  0.55 \pmi 0.07  &  1.33 & 0.29  &  0.32(G) \\ 
18  & 71  &  19\ 43\ 10.33 & 23\ 20\ 51.0  &  1.15 \pmi 0.14 &  0.50 \pmi 0.12  &  1.41 & 0.92  &  ---  \\ 
19  & 70  &  19\ 43\ 10.91 & 23\ 20\ 20.8  &  4.12 \pmi 0.01 &  0.51 \pmi 0.01  &  0.06 & 0.67  &  ---  \\ 
20  & 68  &  19\ 43\ 13.62 & 23\ 19\ 05.8  &  4.60 \pmi 0.02 &  0.51 \pmi 0.01  &  1.11 & 0.44  &  0.99(G) \\ 
21  & 12  &  19\ 43\ 09.60 & 23\ 16\ 35.3  &  0.70 \pmi 0.06 &  0.60 \pmi 0.08  &  0.69 & 1.13  &  0.06(G) \\ 
22  & 11  &  19\ 43\ 12.46 & 23\ 16\ 29.8  &  5.26 \pmi 0.03 &  0.52 \pmi 0.01  &  0.87 & 0.33  &  1.00(G) \\ 
23  & 4   &  19\ 43\ 14.28 & 23\ 17\ 18.5  &  0.55 \pmi 0.01 &  0.54 \pmi 0.01  &  0.01 & 0.05  &  0.08(G) \\ 
24  & 9   &  19\ 43\ 14.69 & 23\ 16\ 01.5  &  4.50 \pmi 0.04 &  0.49 \pmi 0.02  &  1.57 & 0.22  &  0.99(G) \\ 
25  & 8   &  19\ 43\ 15.52 & 23\ 16\ 41.0  &  1.85 \pmi 0.02 &  0.56 \pmi 0.03  &  0.48 & 1.62  &  0.27(G) \\ 
26  & 5   &  19\ 43\ 16.67 & 23\ 17\ 44.7  &  1.67 \pmi 0.15 &  0.52 \pmi 0.14  &  1.42 & 1.18  &  ---  \\ 
27  & 64a &  19\ 43\ 16.98 & 23\ 19\ 48.7  &  3.44 \pmi 0.02 &  0.52 \pmi 0.01  &  0.15 & 0.50  &  ---  \\ 
28  & 65  &  19\ 43\ 18.91 & 23\ 18\ 45.6  &  5.15 \pmi 0.11 &  0.54 \pmi 0.03  &  0.84 & 0.76  &  0.92(G) \\ 
29  & 64  &  19\ 43\ 20.30 & 23\ 19\ 20.6  &  0.85 \pmi 0.02 &  0.52 \pmi 0.03  &  0.24 & 0.18  &  0.11(G) \\ 
30  & 62  &  19\ 43\ 22.55 & 23\ 18\ 26.6  &  1.44 \pmi 0.01 &  0.56 \pmi 0.01  &  0.02 & 0.25  &  1.28(E) \\ 
31  & 61  &  19\ 43\ 21.83 & 23\ 17\ 33.6  &  0.97 \pmi 0.01 &  0.51 \pmi 0.02  &  0.16 & 0.44  &  ---  \\ 
32  & 63  &  19\ 43\ 23.11 & 23\ 18\ 11.6  &  1.46 \pmi 0.03 &  0.54 \pmi 0.03  &  0.23 & 0.41  &  ---  \\ 
33  & 65a &  19\ 43\ 16.90 & 23\ 17\ 26.6  &  1.28 \pmi 0.03 &  0.57 \pmi 0.04  &  0.22 & 0.06  &  ---  \\ 
34  & --- &  19\ 43\ 22.96 & 23\ 15\ 14.7  &  1.07 \pmi 0.09 &  0.56 \pmi 0.13  &  0.88 & 0.52  &  ---  \\ 
35  & 25  &  19\ 43\ 02.04 & 23\ 15\ 49.5  &  3.82 \pmi 0.14 &  0.57 \pmi 0.08  &  1.43 & 0.36  &  0.71(G) \\ 
36  & --- &  19\ 43\ 05.18 & 23\ 17\ 10.7  &  3.08 \pmi 0.01 &  0.50 \pmi 0.01  &  0.06 & 0.52  &  ---  \\ 
37  & 10  &  19\ 43\ 12.04 & 23\ 16\ 10.1  &  6.05 \pmi 0.07 &  0.50 \pmi 0.01  &  0.53 & 0.40  &  ---  \\ 
38  & --- &  19\ 43\ 16.98 & 23\ 16\ 09.5  &  3.61 \pmi 0.02 &  0.52 \pmi 0.01  &  0.16 & 0.74  &  ---  \\ 
39  & --- &  19\ 43\ 14.54 & 23\ 20\ 17.0  &  5.07 \pmi 0.05 &  0.53 \pmi 0.01  &  1.03 & 0.56  &  ---  \\ 
40  & 14  &  19\ 43\ 07.13 & 23\ 16\ 35.8  &  3.18 \pmi 0.12 &  0.55 \pmi 0.06  &  3.59 & 0.22  &  0.72(G) \\  
41  & 33  &  19\ 43\ 16.05 & 23\ 13\ 22.0  &  1.05 \pmi 0.01 &  0.55 \pmi 0.01  &  0.05 & 0.36  &  0.71(S) \\ 
42  & 36  &  19\ 43\ 16.27 & 23\ 11\ 31.7  &  2.42 \pmi 0.01 &  0.56 \pmi 0.01  &  0.01 & 0.74  &  ---  \\ 
43  & --- &  19\ 43\ 14.72 & 23\ 09\ 59.2  &  2.37 \pmi 0.02 &  0.53 \pmi 0.01  &  0.07 & 0.27  &  ---  \\ 
44  & 34  &  19\ 43\ 23.11 & 23\ 12\ 39.5  &  3.70 \pmi 0.01 &  0.54 \pmi 0.01  &  0.33 & 0.22  &  1.07(G) \\ 
45  & 51  &  19\ 43\ 28.70 & 23\ 14\ 38.5  &  2.07 \pmi 0.03 &  0.55 \pmi 0.03  &  0.40 & 1.00  &  ---  \\ 
46  & 53  &  19\ 43\ 32.51 & 23\ 15\ 04.9  &  1.86 \pmi 0.01 &  0.53 \pmi 0.01  &  0.02 & 0.36  &  ---  \\ 
47  & 35  &  19\ 43\ 29.37 & 23\ 10\ 39.7  &  1.59 \pmi 0.02 &  0.53 \pmi 0.02  &  0.62 & 0.22  &  0.79(S) \\ 
48  & 50  &  19\ 43\ 37.73 & 23\ 14\ 02.4  &  0.59 \pmi 0.01 &  0.57 \pmi 0.01  &  0.30 & 0.37  &  ---  \\ 
49  & 49  &  19\ 43\ 38.93 & 23\ 14\ 07.1  &  1.12 \pmi 0.04 &  0.50 \pmi 0.07  &  0.82 & 0.87  &  ---  \\ 
50  & --- &  19\ 43\ 33.47 & 23\ 10\ 01.2  &  1.80 \pmi 0.01 &  0.58 \pmi 0.02  &  0.10 & 0.59  &  ---  \\ 
51  & 42  &  19\ 43\ 44.14 & 23\ 11\ 12.7  &  0.98 \pmi 0.01 &  0.56 \pmi 0.01  &  0.23 & 0.26  &  ---  \\ 
52  & 98  &  19\ 43\ 05.38 & 23\ 24\ 32.8  &  1.47 \pmi 0.02 &  0.59 \pmi 0.02  &  0.11 & 0.16  &  ---  \\ 
53  & 99  &  19\ 43\ 04.96 & 23\ 24\ 17.6  &  2.69 \pmi 0.10 &  0.50 \pmi 0.03  &  0.39 & 0.36  &  ---  \\ 
54  & 80  &  19\ 43\ 05.36 & 23\ 23\ 11.1  &  1.90 \pmi 0.08 &  0.57 \pmi 0.05  &  1.04 & 0.33  &  ---  \\ 
55  & 79  &  19\ 43\ 04.93 & 23\ 22\ 54.4  &  3.21 \pmi 0.01 &  0.55 \pmi 0.01  &  0.02 & 0.57  &  ---  \\ 
56  & --- &  19\ 43\ 09.17 & 23\ 23\ 42.1  &  3.37 \pmi 0.01 &  0.55 \pmi 0.01  &  0.04 & 0.37  &  ---  \\ 
57  & 81  &  19\ 43\ 13.32 & 23\ 22\ 12.2  &  4.41 \pmi 0.22 &  0.51 \pmi 0.07  &  2.22 & 0.74  &  ---  \\ 
58  & --- &  19\ 43\ 12.91 & 23\ 23\ 32.2  &  2.15 \pmi 0.08 &  0.54 \pmi 0.06  &  0.50 & 0.35  &  ---  \\ 
59  & 82  &  19\ 43\ 17.75 & 23\ 22\ 12.3  &  0.91 \pmi 0.01 &  0.52 \pmi 0.01  &  0.09 & 0.09  &  0.92(S) \\ 
60  & 125 &  19\ 43\ 20.03 & 23\ 25\ 42.1  &  0.71 \pmi 0.04 &  0.55 \pmi 0.10  &  1.16 & 0.21  &  ---  \\ 
61  & --- &  19\ 43\ 21.66 & 23\ 24\ 08.7  &  2.77 \pmi 0.14 &  0.60 \pmi 0.08  &  0.98 & 0.35  &  ---  \\ 
62  & --- &  19\ 43\ 23.35 & 23\ 23\ 29.7  &  0.66 \pmi 0.01 &  0.59 \pmi 0.03  &  0.31 & 0.33  &  ---  \\ 
\hline	  
\end{tabular}
\end{minipage}
\label{tab3}
\end{table*}	  
	  
\begin{table*}
\centering
\begin{minipage}{310mm}
\caption{Continuation of Table 5}
\begin{tabular}{llllllllll}
\hline
\hline
ID(M)$^\dag$& ID(B)$^\dag$$^\dag$& R.A(2000J) & DEC(2000J)&{$P_{max}$\pmi $\epsilon$} $(\%)$ & $\lambda_{max}$\pmi$\epsilon$ ($\mu m$)& $\sigma_1$ & $\overline{\epsilon}$& $E(B-V)$ \\
(1)&(2)& (3)&(4)&(5)&(6)&(7)&(8)&(9)\\
\hline
63  &  96  &  19\ 43\ 24.02 & 23\ 23\ 19.2  &  1.74 \pmi 0.04 &  0.51 \pmi 0.03  &  0.20 & 0.33 &  ---  \\ 
64  &  --- &  19\ 43\ 24.83 & 23\ 26\ 14.7  &  3.01 \pmi 0.12 &  0.51 \pmi 0.04  &  0.50 & 0.24 &  ---  \\ 
65  &  --- &  19\ 43\ 26.47 & 23\ 26\ 21.1  &  4.97 \pmi 0.07 &  0.53 \pmi 0.02  &  0.36 & 0.53 &  ---  \\ 
66  &  --- &  19\ 43\ 26.27 & 23\ 20\ 33.3  &  1.65 \pmi 0.02 &  0.51 \pmi 0.02  &  0.14 & 0.40 &  ---  \\ 
67  &  88  &  19\ 43\ 28.09 & 23\ 21\ 13.0  &  0.71 \pmi 0.01 &  0.52 \pmi 0.04  &  0.46 & 0.15 &  ---  \\ 
68  &  --- &  19\ 43\ 30.17 & 23\ 21\ 19.4  &  1.39 \pmi 0.01 &  0.50 \pmi 0.01  &  0.10 & 0.17 &  ---  \\ 
69  &  86  &  19\ 43\ 31.42 & 23\ 20\ 40.3  &  1.11 \pmi 0.11 &  0.51 \pmi 0.11  &  0.76 & 0.11 &  ---  \\ 
70  &  127 &  19\ 43\ 28.34 & 23\ 24\ 34.1  &  3.38 \pmi 0.02 &  0.58 \pmi 0.01  &  1.10 & 1.67 &  ---  \\ 
71  &  130 &  19\ 43\ 30.70 & 23\ 25\ 29.0  &  1.23 \pmi 0.01 &  0.59 \pmi 0.03  &  0.20 & 0.58 &  ---  \\ 
72  &  129 &  19\ 43\ 29.84 & 23\ 25\ 10.7  &  4.09 \pmi 0.07 &  0.50 \pmi 0.02  &  0.44 & 0.56 &  ---  \\ 
73  &  95  &  19\ 43\ 31.27 & 23\ 23\ 08.5  &  1.43 \pmi 0.01 &  0.55 \pmi 0.02  &  0.36 & 0.57 &  0.52(G) \\ 
74  &  93  &  19\ 43\ 32.43 & 23\ 22\ 09.7  &  2.75 \pmi 0.01 &  0.50 \pmi 0.01  &  0.31 & 0.67 &  0.64(G) \\ 
75  &  94  &  19\ 43\ 32.95 & 23\ 22\ 41.6  &  3.01 \pmi 0.13 &  0.51 \pmi 0.04  &  0.43 & 0.83 &  ---  \\ 
76  &  91  &  19\ 43\ 36.55 & 23\ 21\ 08.5  &  3.19 \pmi 0.02 &  0.55 \pmi 0.01  &  0.48 & 0.56 &  0.79(S) \\ 
77  &  --- &  19\ 43\ 40.13 & 23\ 25\ 41.3  &  1.29 \pmi 0.01 &  0.49 \pmi 0.01  &  0.14 & 0.60 &  ---  \\ 
78  &  --- &  19\ 43\ 41.07 & 23\ 24\ 44.0  &  1.59 \pmi 0.08 &  0.49 \pmi 0.04  &  0.48 & 1.17 &  ---  \\ 
79  &  --- &  19\ 43\ 42.50 & 23\ 24\ 42.7  &  2.21 \pmi 0.14 &  0.59 \pmi 0.07  &  0.66 & 0.48 &  ---  \\ 
80  &  --- &  19\ 43\ 11.95 & 23\ 24\ 50.7  &  3.49 \pmi 0.06 &  0.57 \pmi 0.02  &  0.32 & 0.72 &  ---  \\ 
81  &  128 &  19\ 43\ 30.09 & 23\ 24\ 34.4  &  4.25 \pmi 0.06 &  0.56 \pmi 0.02  &  0.52 & 0.24 &  ---  \\ 
82  &  --- &  19\ 43\ 24.58 & 23\ 21\ 15.3  &  0.95 \pmi 0.05 &  0.51 \pmi 0.07  &  0.40 & 0.42 &  ---  \\    	 
83  &  --- &  19\ 42\ 25.70 & 23\ 16\ 16.3  &  1.41 \pmi 0.11 &  0.53 \pmi 0.12  &  1.41 & 0.83 &  ---  \\ 
84  &  --- &  19\ 42\ 28.22 & 23\ 16\ 18.4  &  1.19 \pmi 0.01 &  0.50 \pmi 0.02  &  0.26 & 0.67 &  ---  \\ 
85  &  --- &  19\ 42\ 29.26 & 23\ 13\ 50.4  &  0.77 \pmi 0.01 &  0.55 \pmi 0.02  &  0.66 & 0.75 &  ---  \\ 
86  &  --- &  19\ 42\ 31.16 & 23\ 15\ 00.4  &  2.83 \pmi 0.03 &  0.57 \pmi 0.02  &  0.09 & 0.26 &  ---  \\ 
87  &  --- &  19\ 42\ 32.97 & 23\ 17\ 33.5  &  3.69 \pmi 0.01 &  0.52 \pmi 0.01  &  0.16 & 1.33 &  ---  \\ 
88  &  --- &  19\ 42\ 35.12 & 23\ 16\ 27.4  &  2.46 \pmi 0.01 &  0.53 \pmi 0.01  &  0.02 & 1.25 &  ---  \\ 
89  &  --- &  19\ 42\ 36.55 & 23\ 11\ 22.3  &  3.38 \pmi 0.03 &  0.57 \pmi 0.02  &  0.16 & 0.33 &  ---  \\ 
90  &  --- &  19\ 42\ 38.92 & 23\ 13\ 43.4  &  4.34 \pmi 0.07 &  0.57 \pmi 0.03  &  0.40 & 0.93 &  ---  \\ 
91  &  --- &  19\ 42\ 40.51 & 23\ 14\ 38.1  &  3.26 \pmi 0.02 &  0.52 \pmi 0.02  &  0.25 & 0.47 &  ---  \\ 
92  &  184 &  19\ 42\ 41.38 & 23\ 14\ 41.3  &  3.29 \pmi 0.07 &  0.51 \pmi 0.03  &  0.67 & 0.67 &  ---  \\ 
93  &  179 &  19\ 42\ 42.33 & 23\ 16\ 57.6  &  1.98 \pmi 0.05 &  0.53 \pmi 0.03  &  0.51 & 0.57 &  0.93(E) \\ 
94  &  180 &  19\ 42\ 42.79 & 23\ 16\ 39.5  &  2.57 \pmi 0.01 &  0.52 \pmi 0.01  &  1.10 & 0.22 &  ---  \\ 
95  &  --- &  19\ 42\ 40.72 & 23\ 11\ 18.6  &  2.57 \pmi 0.05 &  0.50 \pmi 0.02  &  0.24 & 0.61 &  ---  \\ 
96  &  191 &  19\ 42\ 45.20 & 23\ 13\ 15.9  &  2.68 \pmi 0.08 &  0.53 \pmi 0.04  &  0.51 & 0.33 &  ---  \\ 
97  &  --- &  19\ 42\ 45.58 & 23\ 14\ 15.0  &  2.28 \pmi 0.01 &  0.57 \pmi 0.03  &  0.16 & 0.76 &  ---  \\ 
98  &  185 &  19\ 42\ 46.19 & 23\ 15\ 02.0  &  2.88 \pmi 0.01 &  0.52 \pmi 0.01  &  0.08 & 0.74 &  ---  \\ 
99  &  22  &  19\ 42\ 50.25 & 23\ 18\ 25.4  &  2.28 \pmi 0.01 &  0.52 \pmi 0.01  &  0.03 & 0.13 &  ---  \\ 
100 &  186 &  19\ 42\ 50.32 & 23\ 15\ 39.1  &  3.77 \pmi 0.01 &  0.55 \pmi 0.01  &  0.32 & 0.67 &  ---  \\ 
101 &  192 &  19\ 42\ 50.17 & 23\ 12\ 45.2  &  2.81 \pmi 0.07 &  0.51 \pmi 0.04  &  0.72 & 1.83 &  ---  \\ 
102 &  187 &  19\ 42\ 56.48 & 23\ 13\ 37.9  &  3.26 \pmi 0.09 &  0.55 \pmi 0.08  &  4.66 & 0.44 &  ---  \\ 
103 &  --- &  19\ 43\ 02.16 & 23\ 16\ 14.1  &  4.22 \pmi 0.01 &  0.51 \pmi 0.01  &  0.02 & 0.21 &  ---  \\ 
104 &  29  &  19\ 43\ 02.13 & 23\ 14\ 30.4  &  4.39 \pmi 0.05 &  0.50 \pmi 0.02  &  1.28 & 0.33 &  ---  \\ 
\hline	  
\end{tabular}
\end{minipage}
\label{tab4}
\begin{quote}
{\hspace{.223cm}{ $^\dag$ : According to this observation}} \\
{\hspace{.16cm}{  $^\dag$ $^\dag$ : According to  Barkhatova (1957) }} \\
{\hspace{.175cm}{(G) : According to Guetter (1992)}}\\
{\hspace{.24cm}{(S) : According to Stone (1988)}}\\
{\hspace{.20cm}{(E) : According to Erickson (1971)}}
\end{quote}
\end{table*}	  

\subsection{Polarization efficiency}

For  interstellar dust  particles in  diffuse interstellar  medium the
ratio between the maximum  amount of polarization to visual extinction
(polarization  efficiency) can  not exceed  the empirical  upper limit
(Hiltner 1956),
\begin{equation}
\hspace*{16mm}{P_{max} < 3A_{V} \simeq 3R_{V} \times E(B-V)}
\end{equation}
The ratio $P_{max}/E(B-V)$ mainly depends on the alignment efficiency,
magnetic strength  and the amount  of depolarization due  to radiation
traversing more than one cloud in different direction.

Figure \ref{eff} shows the relation between colour excess $E(B-V)$ and
maximum  polarization $P_{max}$  for the  stars observed  by  us (open
square  symbol for  non-member  and  star symbol  for  member stars  )
towards NGC 6823  produced by the dust grains along  the line of sight
to  the cluster.   Out of  the 104  stars observed  by us  $E(B-V)$ is
available only  for 26 stars (21  member and 5  non-member stars).  In
the polarization efficiency diagram  three non-member stars \#12, \#16
and \#18  (identification according  to ID(B)) are  lying to  the left
side of  the interstellar maximum  line, which imply that  these three
stars may be affected by  intrinsic polarization. In case of star \#12
the dispersion in the position angle $\overline{\epsilon}$ is found to
be  higher (1.13)  but the  $\sigma_1$  is lower  than the  threshold.
Apparently, the  dominant mechanism  of polarization for  the observed
member stars of NGC 6823 is supposed to be the selective absorption by
the  interstellar dust  grains  which  are aligned  by  the local  and
galactic  magnetic field.   Also the  Figure \ref{eff}  indicates that
while the colour  excess for the member stars of  NGC 6823 varies from
0.27  to 1.28  mag approximately,  the variation  in  the polarization
value  is  very high  $\sim  4.5  \%$.   High variation  of  $P_{max}$
indicates the different  populations of dust grains may  be present in
the line of sight towards NGC 6823, as inferred from the Section 3.1.

In Figure  \ref{eff13}, we plot the $P_{max}/E(B-V)$  vs. $E(B-V)$ for
the   104  stars  observed   by  us   with  available   colour  excess
$E(B-V)$(black open  square symbol for non-member and  star symbol for
member stars ).  The polarization efficiency is found to fall with the
increase in  $E(B-V)$. The  decrement of polarization  efficiency with
increase in  $E(B-V)$ may be  because of increase  in the size  of the
dust grains or small change in the polarization position angle.
 
\section{Summary}

The main results of this study are summarized as follows:

We have observed  the linear polarization for 104  stars in the region
of open  cluster NGC 6823.  The analysis of  these data show  that the
polarization is  mostly due to the foreground  dust grains distributed
in  patchy pattern  and  the majority  of  the observed  stars do  not
present the  indication of intrinsic polarization.  We  also found the
evidence of several dust layers/components  along the line of sight to
the cluster.  Combining our results with those from the literature, we
present the evidence  for the presence of first  layer of dust located
approximately within $200$ pc towards the cluster.

The radial  distribution of the  position angles for the  member stars
are found to show a  systematic change while the polarization found to
reduce towards the  outer parts of the cluster.   The average position
angle of the member stars belongs to the coronal region is more closer
to the  inclination of  the Galactic parallel  ($\sim32^{\circ}$) than
the nuclear region of the cluster.

Polarization  efficiency  for  the  cluster member  stars  are  nearly
similar  to that for  the field  stars, which  imply that  the similar
polarization  mechanism is  responsible for  both the  member  and the
field stars towards the cluster.

The  weighted  mean  of  maximum wavelength  $\lambda_{max}$  for  the
cluster  members and the  field stars  are found  to be  $0.53\pm 0.01
\ \mu  m$ and $0.54\pm  0.01 \ \mu  m$, respectively. These  values of
$\lambda_{max}$ of stars towards NGC 6823 are thus similar to those of
interstellar  medium   ($0.55\pm  0.04   \  \mu  m$).   Therefore  the
polarization towards NGC  6823 is caused mainly due  to the foreground
dust grains as  we have inferred for the clusters IC  1805 and NGC 654
(Medhi  \etal 2007,  2008).  The  intracluster dust  grains  are very
similar to that of the general interstellar medium.

\section*{Acknowledgments}

The  authors  thank  the   referee  Prof.  Carlos  Feinstein  for  his
constructive comments  and suggestions which have  lead to substantial
improvements   of  the  paper.    It  is   also  our   pleasure  thank
Prof. H. C. Bhatt for  his useful suggestions.  This research has made
use of the WEBDA database,  operated at the Institute for Astronomy of
the  University of  Vienna, use  of  image from  the National  Science
Foundation and  Digital Sky  Survey (DSS), which  was produced  at the
Space Telescope  Science Institute under  the US Government  grant NAG
W-2166;  use of  NASA's  Astrophysics  Data System  and  use of  IRAF,
distributed  by  National Optical  Astronomy  Observatories, USA.  The
author (BJM) like to thank  his daughter Sanskriti and wife Orchid for
their support.

\section*{REFERENCES}

Barkhatova K.A. 1957, Soviet Astron. J. 1, 827\\
Chapin, E. L., Ade, P. A. R., Bock, J. J., Brunt, C., Devlin, M. J., Dicker, S. \etal, 2008, ApJ, 681, 428 \\
Clayton, G. C., Cardelli, J. A., 1988, AJ, 96, 695 \\ 
Corradi, Romano L. M., Aznar, R., Mampaso, A., 1998, MNRAS, 297, 617 \\
Coyne, G. V., Magalhaes, A. M., 1979, AJ, 84, 1200 \\
Dame, T. M.; Thaddeus, P.; 1985, Astrophys.J.,170 325.\\ 
Davis, Leverett, Jr.; Greenstein, Jesse L.; 1951, ApJ, 114, 206 \\
Dias, W. S., Assafin, M., Flório, V., Alessi, B. S., Líbero, V.,2006, A\&A, 446, 949 \\
Dobashi, K., Uehara, H., Kandori, R., Sakurai, T., Kaiden, M., Umemoto, T., Sato, F., 2005, PASJ, 57, 1 \\
Erickson R.R., 1971, A\&A,10, 270 \\ 
Feinstein, A., 1994, RMxAA, 29, 141 \\
Fresneau, A.; Monier, R.; 1999, AJ,118,421.\\
Goldreich, Peter, Kwan, John, 1974, ApJ, 189, 441 \\
Guarinos J., 1992, Astronomy from Large Databases II", ESO Conference and Workshop Proceedings No 43, ISBN 3-923524-47-1, 301 \\
Guetter, H. H.; 1992, AJ 103, 197. \\
Guetter, Harry H., Vrba, Frederick J., 1989, AJ, 98, 611 \\
Hall, John Scoville, 1958, PUSNO, 17, 1 \\ 
Heiles, Carl; 2000, AJ, 119, 923 \\
Hiltner, W. A., 1956, ApJS, 2, 389 \\
Hoag, A. A., Johnson, H. L., Iriarte, B., Mitchell, R. I., Hallam, K. L., Sharpless, S.; 1961, PUSNO, 17, 343 \\
H$\phi$g, E., B\"{a}ssgen, G., Bastian, U., Egret, D., Fabricius, C., Gro$\beta$mann, V. \etal, 1997, A\&A, 323, 57 \\
Jones, R. Victor, Spitzer, Lyman, Jr., 1967, ApJ, 147, 943 \\
Kharchenko N.V., Piskunov A.E., Röser S., Schilbach E., Scholz R.D., 2005, A\&A, 438, 1163 \\
Lazarian, A., Goodman, Alyssa A., Myers, Philip C., 1997, ApJ, 490, 273 \\
Leisawitz, David, Bash, Frank N., Thaddeus, Patrick, 1989, ApJS, 70, 731 \\
Lynds, Beverly T., 1962, ApJS, 7, 1 \\
McMillan, R. S., 1978, APJ, 225, 880 \\
Medhi, Biman J., Maheswar, G., Brijesh, K., Pandey, J. C., Kumar, T. S., Sagar, R.,2007, MNRAS, 378, 881 \\
Medhi, Biman J., Maheswar, Pandey, J. C., Kumar, T. S., Sagar, R.,2008, MNRAS, 388, 105 \\
Morgan, W. W., Whitford, A. E. and Code, A. D.; 1953, Astrophys.J.118,318.\\
Neckel, Th., Klare, G., 1980, A\&ASS, 42 251. \\
Orsatti, A. M., Vega, E., Marraco, H. G., 1998, AJ, 116, 266 \\	
Pandey, J. C., Medhi, Biman J., Sagar, R.,  Pandey, A. K., 2009,  MNRAS, 396, 1004 \\  
Perryman, M. A. C., Lindegren, L., Kovalevsky, J., H$\phi$g, E., Bastian, U., Bernacca, P. L. \etal, 1997, A\&A, 323, 49 \\
Ramaprakash, A. N., Gupta, R., Sen, A. K., Tandon, S. N., 1998, A\&AS, 128, 369 \\
Rautela, B. S., Joshi, G. C., Pandey, J. C., 2004, BASI, 32,159 \\
Sagar, R. and Joshi, U. C.; 1981, Ap\&SS 75, 465. \\
Sagar, R., Joshi, U. C., Sinvhal, S. D., 1983, BASI, 11, 44 \\
Sagar, R.; 1987, MNRAS, 228, 483 \\
Serkowski, K., 1965, ApJ, 141, 1340 \\
Serkowski, K., 1973, IAUS, 52, 145 \\
Serkowski, K.; Mathewson, D. L.; Ford, V. L.;  1975, ApJ, 196, 261 \\
Schmidt, G. D., Elston, R., Lupie, O. L., 1992, AJ, 104, 1563 \\
Spitzer, Lyman, 1978, Physical processes in the interstellar medium, John Willy \& sons, New York \\
Stone, Ronald C.,1988, AJ, 96, 1389 \\
Turner, D. G.,  1979, JRASC, 73, 74 \\
Turnshek, D. A., Bohlin, R. C., Williamson, R. L., II, Lupie, O. L., Koornneef, J., Morgan, D. H., 1990, AJ, 99, 1243 \\
Wilking, B. A., Lebofsky, M. J., Kemp, J. C., Martin, P. G., Rieke, G. H., 1980, ApJ, 235, 905 \\
Whittet, D.C.B., van Breda, I.G., 1978, A\&A, 66, 57 \\
Whittet, D. C. B., 1992, Dust in the Galactic Environment (Bristol:IOP)\\	  

\bsp
\label{lastpage}
\end{document}